\documentclass[11pt]{article}
\usepackage{a4wide}
\usepackage{amstext,amsfonts,amsmath}

\begin{document}
\begin{flushright}
FTUV--04-0908  \quad IFIC--04--50\\
September 8, 2004\\
Ann. Phys. {\bf 317}, 238-279 (2005)
\end{flushright}
\thispagestyle{empty}

\vspace{1cm}

\begin{center}
{\LARGE\bf On the formulation of $D$=11 supergravity and }
\vspace{0.5cm} {\LARGE\bf the composite nature of its three--form
gauge field}

\vspace{2cm}

{\large\bf Igor A. Bandos$^{\dagger,\ast}$, Jos\'e A. de
Azc\'arraga$^{\dagger}$,
 Mois\'es Pic\'on$^{\dagger}$ and Oscar
Varela$^{\dagger}$}

\vspace{1cm}

{\it $^{\dagger}$Departamento de F\'{\i}sica Te\'orica, Universidad de
Valencia and IFIC (CSIC-UVEG), 46100-Burjassot (Valencia), Spain
\\
$^{\ast}$Institute for Theoretical Physics, NSC ``Kharkov Institute of
Physics  and Technology'',  UA61108, Kharkov, Ukraine}

\end{center}
\def\theequation{\arabic{section}.\arabic{equation}}

\vspace{1.5cm}

\begin{abstract}
The underlying gauge group structure of the $D=11$
Cremmer-Julia-Scherk supergravity becomes manifest when its
three-form field $A_3$ is expressed through a set of one--form
gauge fields, $B_1^{a_1 a_2}$, $B_1^{a_1 \ldots a_5}$,
$\eta_{1\alpha}$ and $E^a$, $\psi^\alpha$. These are associated
with the generators of the elements of a family of enlarged
supersymmetry algebras $\tilde{\mathfrak{E}}^{(528|32+32)}(s)$
parametrized by a real number $s$. We study in detail the
composite structure of $A_3$ extending previous results by D'Auria
and Fr\'e, stress the equivalence of the above problem to the
trivialization of a standard supersymmetry algebra
${\mathfrak{E}}^{(11|32)}$ cohomology four-cocycle on the enlarged
${\tilde {\mathfrak{E}}}^{(528|32+32)}(s)$ superalgebras, and
discuss its possible dynamical consequences. To this aim we
consider the properties of the first order supergravity action
with a composite $A_3$ field and find the set of extra gauge
symmetries that guarantee that the field theoretical degrees of
freedom of the theory remain the same as with a fundamental $A_3$.
The extra gauge symmetries are also present in the so--called
rheonomic treatment of the first order $D=11$ supergravity action
when $A_3$ is composite. Our considerations on the composite
structure of $A_3$ provide one more application of the idea that
there exists an extended superspace coordinates/fields
correspondence. They also suggest that there is a possible
embedding of $D=11$ supergravity into a theory defined on the
enlarged superspace ${\tilde{\Sigma}}^{(528|32+32)}(s)$.

\end{abstract}

\vspace{1cm}

\noindent


\newpage

\setcounter{page}1

\section{Introduction}

Already in the original paper \cite{CJS} where the standard $D=11$
supergravity theory was introduced, Cremmer, Julia and Scherk
(CJS) considered its possible association with a gauge theory and
suggested that the gauge group could be related to $OSp(1|32)$.
However, the explicit form of such a connection was unclear as
{\it e.g.}, an \.In\"on\"u--Wigner contraction of $OSp(1|32)$ did
not allow for the spin connection among the set of gauge fields.
Also the relation of the three-form gauge field $A_3= {1\over 3!}
dx^\mu\wedge dx^\nu \wedge dx^\rho A_{\rho\nu\mu}(x)$ with such a
Lie superalgebra was unclear \footnote{See \cite{Dualdual2} for an
interesting treatment of $p$-form gauge fields as Goldstone fields
of Lie superalgebras, \cite{BNS03} for the corresponding sigma
model-like action for supergravity and \cite{Kallosh84} for a
reformulation of 11--dimensional supergravity as a theory of the
graviton, the gravitino and an independent spin connection in
which $A_3$ is treated as a composite of $E^a$, $\psi^\alpha$ and
$\omega^{ab}$.}. The problem was addressed in \cite{D'A+F} where
it was found, in particular, that the three-form $A_3$ of CJS
supergravity \cite{CJS} can be expressed through the graviton
$E^a(x)=dx^\mu e_\mu^a(x)$, the gravitino $\psi^\alpha (x)= dx^\mu
\psi^\alpha_\mu(x)$, the additional  bosonic one--forms
$B_1^{ab}(x)= dx^\mu B_\mu^{ab}(x)$, $B_1^{a_1\ldots a_5} (x)=
dx^\mu B_\mu^{a_1\ldots a_5}(x)$, and an additional fermionic
one-form $\eta_{1 \alpha}(x)= dx^\mu \eta_{\mu\,\alpha}(x)$.

Although the presence of additional fermionic fields is
undesirable for the standard eleven-dimensional supersymmetry, the
presence of the $\eta_{\mu \alpha}$ field is not a problem in the
context of \cite{D'A+F}. First, it corresponds to central
fermionic generators. Secondly, the additional gravitino-like
field $\eta_{\mu\alpha}$ appears in the description of  $D=11$
supergravity only through  the three-form field $A_3$, which is
considered as a composite of the `old' ($E^a$, $\psi^\alpha$) and
`new' ($B_1^{ab}, B_1^{abcde}, \eta_{1 \alpha}$) fields,
\begin{eqnarray}
\label{A3=A3def} A_3 = A_3( E^a \,, \, \psi^\alpha \; ; \;
B_1^{ab}, B_1^{abcde}, \eta_{1 \alpha} \,)\;;
\end{eqnarray}
the new bosonic fields also appear through $A_3$ only.

The composite  structure in Eq.~(\ref{A3=A3def}) suggests
\cite{D'A+F} a possible underlying gauge symmetry of the $D=11$
supergravity. The new fields $B_\mu^{ab}(x)$, $B_\mu^{abcde}(x)$,
$\eta_{\mu \alpha}(x)$ may be treated as gauge fields associated
with new antisymmetric tensor generators $Z_{ab}=Z_{[ab]}$,
$Z_{abcde}= Z_{[abcde]}$ and a new fermionic generator
$Q^{\prime\alpha}$, which extends  the super--Poincar\'e algebra
in which $Q_\alpha$, $P_a$ and $M_{ab}$ correspond to the
gravitino field $\psi^\alpha = dx^\mu \psi^\alpha_\mu(x)$, the
graviton $E^a(x)=dx^\mu e_\mu^a(x)$  and the spin connection
$\omega^{ab}= dx^\mu \omega_\mu^{ab}(x)$. The possibility of
constructing $A_{\mu\nu\rho}$ from the above set of gauge fields
fixes the free differential algebra of $B_1^{ab}(x)$,
$B_1^{abcde}(x)$, $\eta_{1 \alpha}(x)$ and, hence, the algebra of
generators $P_a$, $M_{ab}$, $Q_\alpha\, $, $\; Z_{ab}$,
$Z_{abcde}$, $Q^{\prime\alpha}$ \cite{D'A+F} (see below and
Sec.~\ref{Sectriv} for details). Two possible superalgebras
allowing for a composite nature of $A_3$ were found in
\cite{D'A+F}; we will call them `D'Auria--Fr\'e superalgebras'.
Both of them are central extensions of the M-theory superalgebra
or M-algebra \cite{M-alg} (see also \cite{HP82}) \footnote{See
\cite{gM-alg,JdA00} and refs. therein for further generalizations
of the M--theory superalgebra and for their structure.},
\begin{eqnarray}\label{M-alg}
\{ Q_\alpha , Q_\beta \} & = & P_{\alpha \beta} \; , \qquad
{} [ P_{\alpha\beta} , P_{\gamma\delta} ] =  0 \; , \qquad
\nonumber \\
 P_{\alpha\beta} =  P_{\beta\alpha} &=& P_a \Gamma^a_{\alpha\beta} +
Z_{ab} i\Gamma^{ab}_{\alpha\beta} +  Z_{a_1\ldots a_5}
\Gamma^{a_1\ldots a_5 }_{\alpha\beta} \; , \qquad
\\ \label{[QP]=0}
 \left[ Q_\alpha , P_{\beta\gamma} \right] & = & 0 \; ,  \quad
\end{eqnarray}
by a new fermionic central charge $Q^{\prime\alpha}$. These
algebras are  defined by Eqs. (\ref{M-alg}) plus
\begin{eqnarray}\label{[Z,Q]}
{} [ P_{a} , Q_\alpha ] & = & \delta (\Gamma_a Q^\prime)_\alpha \,
, \qquad \nonumber
\\ {} [
Z_{ab} , Q_\alpha ] & = &   i\gamma_1 (\Gamma_{ab}
Q^\prime)_\alpha \; , \qquad \nonumber
\\  {} [ Z_{abcde} , Q_\alpha ] & = & \gamma_2 (\Gamma_{abcde}
Q^\prime)_\alpha \; ,  \qquad \nonumber
\\ \label{[Q',]} {}
[Q^{\prime\alpha} \!\! & , &\! all\; \}  =  0 \; ,
\end{eqnarray}
for two sets of specific values of the constants $\delta, \gamma_1,
\gamma_2$. In general,  Eqs. (\ref{M-alg}), (\ref{[Z,Q]}) define a
one--parametric family of superalgebras, since the allowed values of
constants $\delta, \gamma_1, \gamma_2$ are restricted \cite{D'A+F}
only by the Jacobi identity
\begin{equation}
\label{idg0} \delta + 10 \gamma_1- 6! \gamma_2=0 \; ;
\end{equation}
one parameter, $\gamma_1$ if nonzero and $\delta$ otherwise, may be
absorbed in the normalization of the central fermionic generator
$Q^{\prime\alpha}$ and (in this sense) is inessential.

The essential parameter $s$ distinguishing the non-isomorphic
members of the
 family $\tilde{\mathfrak{E}}(s)=\tilde{\mathfrak{E}}^{(528|32+32)}(s)$
\cite{Lett}\footnote{We shall denote by ${\mathfrak{E}}$
($\tilde{{\mathfrak{E}}}$) the supersymmetry (enlarged
supersymmetry) algebras associated with the corresponding rigid
superspace supergroups, denoted $\Sigma$ (${\tilde{\Sigma}}$). The
symbols $\Sigma$, ${\tilde{\Sigma}}$ will also be used to denote
the corresponding non--flat superspaces (in which case there is no
group structure) without risk of confusion.} of the fermionic
central extensions (hence $32+32$ and not just $64$) of the
M-algebra can be introduced {\it e.g.} by parametrizing $\delta,
\gamma_1, \gamma_2$ as follows:
 \begin{eqnarray}
\label{s-def} s:= {\delta \over 2\gamma_1} - 1
\quad\;  & \Rightarrow & \quad \left\{
{\setlength\arraycolsep{2pt}\begin{array}{lll} \delta
&=& 2\gamma_1(s+1) \, , \\
\gamma_2 &=& 2\gamma_1({s \over 6!} + {1 \over 5!}) \; .
\end{array}} \right.    \;
\end{eqnarray}
(this makes sense for $\gamma_1\neq 0$; to apply this for $\gamma_1
=0$ one should consider the limit  $\gamma_1 \rightarrow 0$,
$s\rightarrow \infty$, $\gamma_1 s = \delta/2 \rightarrow finite$).
The properties of the two specific D'Auria--Fr\'e superalgebras
($\tilde{\mathfrak{E}}(3/2)$ and $\tilde{\mathfrak{E}}(-1)$  in the
above notation) did not have a clear origin. This question was taken
up in \cite{CFGPN} and, in particular, whether these two
superalgebras could be contractions of $osp(1|64)$ or $su(32|1)$.
The answer was negative, and the authors of \cite{CFGPN} noted the
possibility of looking at non-semisimple supergroups involving
$OSp(1|32)$ in such a context.

Recently we have found \cite{Lett} that {\it all the $s\not=0$
members} of the family $\tilde{\mathfrak{E}}(s)$ allow for a
composite  $A_3$ expressed in terms of one--form gauge fields. This
implies that $D$=11 supergravity possesses a gauge symmetry under
the $\tilde{\Sigma}(s\not=0 )\times\!\!\!\!\!\! \supset SO(1,10) =
\tilde{\Sigma}^{(528|32+32)}(s\not=0)\times\!\!\!\!\!\!\supset
SO(1,10)$ supergroup associated with the
$\tilde{\mathfrak{E}}(s\not=0 )+\!\!\!\!\!\!\supset so(1,10)$ Lie
superalgebra. This {\it underlying} gauge symmetry is hidden in the
original CJS formulation but becomes manifest in the $D$=11
supergravity with a composite $A_3$.

 Although the presence of a family of superalgebras
$\tilde{\mathfrak{E}}(s\not=0 )+\!\!\!\!\!\!\supset so(1,10)$,
rather than a unique one, may  indicate that the found answer on
the hidden gauge group structure of the $D=11$ supergravity is not
the final one, the origin of these hidden symmetry supergroups is
now clearer \cite{Lett}. Firstly, all the corresponding
supergroups $\tilde{\Sigma}(s\not=0 )\times\!\!\!\!\!\!\supset
SO(1,10)$ are nontrivial deformations of
$\tilde{\Sigma}(0)\times\!\!\!\!\!\!\supset SO(1,10)$, and the
latter, as well as $\tilde{\Sigma}(0)\times\!\!\!\!\!\!\supset
Sp(32)\supset \tilde{\Sigma}(0)\times\!\!\!\!\!\!\supset SO(1,10)$
are {\it expansions}\footnote{The {\it expansion method}
\cite{H01,JdA02} is a new method of generating new Lie algebras
starting from a given algebra. It includes \cite{JdA02} the
\.In\"on\"u--Wigner and generalized contractions as a particular
case, but in general leads to algebras of larger dimension than
the original one.} of the $OSp(1|32)$ supergroup \cite{Lett}.

In this paper we give further details of the derivation of the
above results on the composite structure of the $A_3$
three-(super)form {\it i.e.}, of the hidden gauge symmetry of
$D=11$ supergravity under (any of) the $\tilde{\Sigma}(s\not=0
)\times\!\!\!\!\!\! \supset SO(1,10)$ supergroups, and study some
of its  possible dynamical consequences. To this end we consider
the spacetime (component) approach, the standard superspace one
and the intermediate rheonomic \cite{Regge,D'A+F,rheoB} approach
to CJS $D=11$ supergravity when the $A_3$ (super)field is
composite. To this aim, we consider
 the original proposal \cite{D'A+F} of substituting Eq.
(\ref{A3=A3def}) for $A_3$ into the first--order formulation of
$D=11$ supergravity action (also proposed in \cite{D'A+F}, see
also \cite{J+S99}\footnote{The action of \cite{D'A+F,J+S99} is
first order both for the gravity part and for the $A_3$ field. An
action that is first order with respect to the gravitational part
but second order in $A_3$  was constructed in \cite{CFGPN}; a
first order formulation in $A_3$ but with a composite spin
connection was given in \cite{PvanN99}.}). We find that such a
dynamical system presents `extra' gauge symmetries [denoted
`extra' to distinguish them from those associated with
$\tilde{\Sigma}(s\not=0)\times\!\!\!\!\!\!\supset SO(1,10)$].
These make the number of degrees of freedom with a composite $A_3$
the same as those of the standard CJS supergravity \cite{CJS}.

The extra gauge symmetries resulting from the composite structure
of $A_3$ are also present in the `rheonomic' action
\cite{Regge,rheoB} for $D$=11 supergravity \cite{D'A+F}. This is
given by the first order action \cite{D'A+F,J+S99} where all the
fields are replaced by superfields and the integration surface is
an arbitrary bosonic surface ${\cal M}^{11}$ in standard
superspace $\Sigma^{(11|32)}$. This composite structure of $A_3$
makes natural to consider ${\cal M}^{11}$ in the reonomic action
as a surface in the enlarged superspace $\tilde{\Sigma}(s)=
\tilde{\Sigma}^{(528|32+32)}(s)$. This suggests an embedding of
$D=11$ supergravity into a theory in a $D=11$ enlarged superspace
$\tilde{\Sigma}^{(528|32+32)}(s\not=0)$. This is supported by
observing that, as we stress in this paper, the search for a
composite structure for the $A_3$ field along \cite{D'A+F,Lett} is
equivalent to solving the problem of trivializing  a
Chevalley--Eilenberg \cite{CE48}  (CE) four-cocycle of the
standard supersymmetry algebra ${\mathfrak{E}}=
{\mathfrak{E}}^{(11|32)}$ cohomology. This requires moving from
${\mathfrak{E}}^{(11|32)}$ to
$\tilde{{\mathfrak{E}}}=\tilde{{\mathfrak{E}}}^{(528|32+32)}(s\not=0)$,
the supersymmetry algebra of the rigid enlarged superspace
$\tilde{\Sigma}^{(528|32+32)}(s\not=0)$. In this perspective the
composite character of the $A_3$ field, {\it i.e.}, the fact that
it may be written in terms of one--form fields associated with a
larger supersymmetry group, can be considered as a further example
of the extended superspace coordinates/(super)fields
correspondence\footnote{The idea of field-space democracy is
explicitly stated in Berezin \cite{Be-79} (`supermathematics...
contains a hint about the existence of a fundamental symmetry
between coordinates and fields') and is implicit in the work of D.
V. Volkov \cite{Volkov73}. The field-space {\it democracy}
framework was further discussed in \cite{Sch-et-al-80} in the
context of the Ogievetski--Sokatchev formulation of $D=4$ $N=1$
superfield supergravity. The case for a (worldvolume)
fields/extended superspace coordinates {\it correspondence}
principle for superbranes has been advocated in \cite{JdA00} (see
also \cite{Az-Iz-Mi-04} in the context of $\kappa$-symmetry).}.

The paper is organized as follows. In Sec. 2 we present a brief
review of the standard superfield (Sec. 2.2), spacetime component
(Sec. 2.1, 2.4) and rheonomic (Sec. 2.5) approaches to $D=11$ CJS
supergravity. We point out the r\^ole  of free differential
algebras (FDAs) in the supergravity description (Sec. 2.3),
describe their relation with {\it Lie} superalgebras and enlarged
superspaces and stress, in this perspective, the peculiarity of
$D=11$ supergravity due to the presence of the three--form
field\footnote{Notice that other higher dimensional supergravities
also include higher form fields. For instance, $D$=10 type IIB
supergravity includes the RR (Ramond--Ramond) four--form $C_4$ and
two two--form gauge fields, the NS--NS
(Neveu-Schwarz---Neveu-Schwarz) two--form $B_2$ and the RR one
$C_2$. Thus, our discussions on enlarged superspaces and hidden
gauge symmetries are relevant there too.\\
${\quad}$ Notice also the appearance of 32 fermionic additional
coordinates
 (associated with the Green algebra) in a recent analysis of
 covariant superstring quantization \cite{GPN-03}. } $A_3$. As we discuss
in Sec. 3, $A_3$ cannot be associated with a Maurer--Cartan (MC)
form of a Lie algebra; rather,  $dA_3$ is associated with a
nontrivial CE four--cocycle of the ${\mathfrak{E}}^{(11|32)}$
cohomology. In Sec. 4 we give the details of the derivation of our
recent result \cite{Lett} on the expression of $A_3$ in terms of
the one--form gauge fields of a one--parametric family of
superalgebras, which are nontrivial deformations of an {\it
expansion} of the $osp(1|32)$ superalgebra denoted
$osp(1|32)(2,3,2)$ (see \cite{JdA02} for the notation). We stress
the equivalence of this problem to that of trivializing the
${\mathfrak{E}}^{(11|32)}$ CE four--cocycle on the extended
algebra ${\tilde {\mathfrak{E}}}^{(528|32+32)}(s)$, and describe
how our family of composite $A_3$ structures includes the two
D'Auria and Fr\'e ones as particular cases. Another member of our
family gives a particularly simple form of $A_3$ that does not
involve a five-index one-form gauge field. In Sec. 5 we study the
consequences of the composite structure of $A_3$ for the first
order supergravity action (Sec. 5.1)  and find a set of extra
gauge symmetries which reduces the number of degrees of freedom to
those of  the action with a fundamental or `elementary' $A_3$
field. These extra gauge symmetries are also shown to be present
in the rheonomic action (Sec. 5.3) with a composite $A_3$, which
then can be treated as an integral over an eleven--dimensional
bosonic surface in the enlarged superspace
$\tilde{\Sigma}^{(528|32+32)}(s)$. This suggests an embedding of
$D$=11 supergravity in a theory defined on such an enlarged
superspace. Our conclusions are presented in Sec.~6.

\setcounter{equation}0
\section{Free differential algebras, superspace constraints and first order
action of $D=11$ supergravity}

\subsection{Differential forms in $D=11$ supergravity}

Any formulation of CJS supergravity involves the graviton,
$e_\mu^a(x)$, the gravitino $\psi_\mu^\alpha(x)$, and the
antisymmetric tensor field $A_{\mu_1\mu_2\mu_3}(x)$, as well as the
spin connection $\omega_\mu^{ab}(x)$. This last one is considered to
be a composite of physical fields (in the second order approach) or
becomes composite on the mass shell (in the first order approach,
see \cite{D'A+F,CFGPN,rheoB,J+S99}). All these fields may be
associated with a set of  differential forms on $D=11$ spacetime
$M^{11}$
\begin{eqnarray}\label{M11dfp}
&& E^a(x) =dx^\mu e_\mu^a(x)\; , \qquad \psi^\alpha(x) =
dx^\mu \psi_\mu^\alpha(x) \; , \qquad  \nonumber \\
&& A_3(x)= {1\over 3!} dx^{\mu_1} \wedge dx^{\mu_2} \wedge
dx^{\mu_3} A_{\mu_3\mu_2\mu_1}(x) \; , \nonumber \\
\label{M11dfa} && \omega^{ab}(x)= dx^\mu \omega_\mu^{ab}(x)\; .
\qquad
\end{eqnarray}
Further one may introduce the gauge field
\begin{eqnarray}
&&  A_6(x)= {1\over 6!} dx^{\mu_1} \wedge \ldots \wedge dx^{\mu_6}
A_{\mu_6\ldots\mu_1}(x) \; ;
\end{eqnarray}
its field strength $F_7(x)= dA_6 +A_3\wedge dA_3$  is dual to the
field strength $F_4(x)=dA_3$ of $A_3(x)$, $\; F_7(x)= \ast F_4(x)$
(see Eq. (\ref{F=*F}) below).

\subsection{On--shell superspace constraints for  $D=11$ supergravity}
\label{sugraconstr}

The above fields may also be associated with a set of superforms
on the standard $D=11$ superspace $\Sigma^{(11|32)}$ with
coordinates $Z^M= (x^\mu , \theta^{\check{\alpha}})$,
 \begin{eqnarray}\label{S11Ebf}
E^a(Z)&=& dZ^M E_M^a(Z)\; , \qquad   \nonumber \\
\psi^\alpha(Z) &:=&  E^\alpha(Z)=
dZ^M E_M^\alpha(Z) \; , \qquad \\
\label{S11EA} &&  E^A:=( E^a, E^\alpha)\; ,   \\
\label{S11A3} A_3(Z) \!&=&\! {1\over 3!} dZ^{M_1} \wedge dZ^{M_2}
\wedge dZ^{M_3} A_{M_3M_2M_1}(Z) \equiv {1\over 3!} E^{A_1}
\wedge E^{A_2} \wedge E^{A_3} A_{A_3A_2A_1}(Z) \, , \quad \\
\label{S11dfa}
 \omega^{ab}(Z) &=& dZ^M \omega_M^{ab}(Z)\equiv E^C \omega_C^{ab}(Z)
\; , \qquad \\ \label{S11dfa6} A_6(Z) &=& {1\over 6!} dZ^{M_1}
\wedge \ldots \wedge dZ^{M_6} A_{M_6\ldots M_1}(Z) \; ,
\end{eqnarray}
 provided these superform potentials obey the  superspace supergravity
constraints \cite{CremmerFerrara80,BrinkHowe80,Lechner93}
\begin{eqnarray}\label{Ta=}
T^a&=&-  i E^\alpha \wedge E^\beta \Gamma^a_{\alpha\beta} \; ,
\\ \label{Tf=}
T^\alpha&=&- {i \over 18} E^a \wedge E^\beta
\left(F_{ac_1c_2c_3}\Gamma^{c_1c_2c_3} + {1\over 8}
F^{c_1c_2c_3c_4} \Gamma_{a c_1c_2c_3c_4}\right)_\beta^{\;\alpha} +
{1 \over 2} E^a \wedge E^b T_{ba}{}^\alpha(Z) \; , \qquad \\
\label{RL=} R^{ab}&=&E^\alpha \wedge E^\beta \left( -{1\over3}
F^{abc_1c_2}\Gamma_{c_1c_2} + {i\over 3^. 5!} (\ast
F)^{abc_1\ldots c_5} \Gamma_{c_1\ldots c_5} \right)_{\alpha\beta}
+ \nonumber
\\ &&   \qquad +  E^c \wedge E^\alpha \left(
-iT^{ab\beta}\Gamma_{c}{}_{\beta\alpha} + 2i T_c{}^{[a \, \beta}
\Gamma^{b]}{}_{\beta\alpha} \right) + {1 \over 2} E^d \wedge E^c
R_{cd}{}^{ab}(Z) \; ,
\\ \label{cF4=}
{\cal F}_4 &:=& dA_3={1\over 2} E^\alpha \wedge E^\beta \wedge
\bar{\Gamma}^{(2)}_{\alpha\beta} + {1 \over 4! } E^{c_4}
\wedge \ldots \wedge E^{c_1} F_{c_1\ldots c_4}(Z) \; , \\
\label{cF7=}  {\cal F}_7 &:=& dA_6 + A_3\wedge dA_3={i\over 2}
E^\alpha \wedge E^\beta \wedge \bar{\Gamma}^{(5)}_{\alpha\beta} +
{1 \over 7! } E^{c_7} \wedge \ldots \wedge E^{c_1} F_{c_1\ldots
c_7}(Z)  \; .
\end{eqnarray}
In the above Eqs. (\ref{Ta=})--(\ref{cF7=}) $T^a$, $T^\alpha$,
$R^{ab}$ are the torsion and curvature two-forms,
\begin{eqnarray}\label{Ta=def}
T^a &:=& DE^a(Z) := dE^a - E^b\wedge \omega_b{}^a \; ,
\\ \label{Tf=def}
T^\alpha &=& DE^\alpha(Z) := dE^\alpha -
E^\beta \wedge \omega_\beta{}^\alpha  \; ,
\\
\label{RL=def} R^{ab} &:=& d\omega^{ab} - \omega^{ac}\wedge
\omega_c{}^b \; ,
\end{eqnarray}
$\omega^{ab}$ is the spin connection,
\begin{eqnarray}
\label{omff} \omega_\beta{}^\alpha &:= & {1\over 4}
\omega^{ab}\Gamma_{ab}{}_\beta{}^\alpha \; ,
\end{eqnarray}
${\cal F}_4 =dA_3$ and ${\cal F}_7 =dA_6 + A_3\wedge dA_3$ are
the  field strength superforms, and we have used the notation
\begin{eqnarray}
\label{Gammaq} \bar{\Gamma}^{(2)}_{\alpha\beta} &:=& {1\over 2}
E^b \wedge E^a
\Gamma_{ab}{}_{\alpha\beta} \; , \nonumber \\
\bar{\Gamma}^{(5)}_{\alpha\beta} &:=& {1\over 5!} E^{a_5} \wedge
\ldots \wedge E^{a_1} \Gamma_{a_1 \ldots a_5} {}_{\alpha\beta} \;
{}.
\end{eqnarray}

As discussed in \cite{CremmerFerrara80,BrinkHowe80,Lechner93}, the
study of the Bianchi identities
 \begin{eqnarray}\label{BITa}
&& DT^a  \equiv -  E^b \wedge R_b{}^a \; ,
 \\
\label{BITal} && DT^\alpha \equiv - E^\beta \wedge
R_\beta{}^\alpha := - {1\over 4} E^\beta \wedge R^{ab} \;
\Gamma_{ab}{}_\beta{}^\alpha \; ,
 \\
\label{DR} && DR^{ab}\equiv 0 \; ,
 \\
\label{DF4}
 &&   d{\cal F}_4 \equiv 0  \; ,
 \qquad   \\
\label{DF7} && d{\cal F}_7 - {\cal F}_4 \wedge {\cal F}_4 \equiv 0
\; ,
\end{eqnarray}
shows that the set of constraints (\ref{Ta=})--(\ref{cF7=}) is
consistent provided that the Riemann tensor $R_{cd}{}^{ab}$ and
the field strengths of the gravitino ($T_{ab}{}^\alpha$) and  of
the gauge field ($F_{c_1\ldots c_4}(Z)$)
obey the (superfield generalizations of the) equations of motion.

Actually, the system of the constraints (\ref{Ta=})--(\ref{cF7=})
is over-complete. This is indicated by the fact that the gauge
field strengths $F_{c_1\ldots c_4}(Z)$ already enter in the
expressions for the torsion (\ref{Tf=}) and the curvature
(\ref{RL=}) of superspace. Indeed, the torsion constraints
(\ref{Ta=}), (\ref{Tf=}) and (\ref{RL=}) already imply the above
mentioned dynamical equations and provide the automatic
consistency of the remaining constraints (\ref{cF4=}),
(\ref{cF7=}), as may be seen by  studying the Bianchi indentities
(\ref{BITa}), (\ref{BITal}), (\ref{DR}) (see \cite{Howe97} for an
even stronger result). However, when the differential superforms
$A_3$ and $A_6$ are introduced, the study of the Bianchi
identities simplifies essentially, which provides a shortcut that
was already used in the first papers
\cite{CremmerFerrara80,BrinkHowe80}. For instance, studying the
Bianchi identities (\ref{DF4}) with the constraints (\ref{Ta=}),
(\ref{cF4=}) and $T^\alpha =  E^b \wedge E^\beta \; T_{\beta
b}{}^\alpha + {1 \over 2} E^a \wedge E^b \; T_{ba}{}^\alpha(Z)$
instead of the specific form of (\ref{Tf=}), one finds, in
addition to $D_{[c_5} F_{c_1c_2c_3c_4]}=0$ (indicating that
$F_{c_1c_2c_3c_4}$ is the field strength of a $A_{c_1c_2c_3}$) and
$D_{\alpha}F_{c_1c_2c_3c_4} = - 3! T_{[c_1c_2}{}^\beta
\Gamma_{c_3c_4]}{} _{\beta }{}^\alpha$, the equation $T_{(\beta|
[a}{}^\gamma \Gamma_{bc]\; |\alpha) \gamma} = i/3! \, F_{abcd}\,
\Gamma^d_{\alpha\beta}$. The solution of this equation expresses
$T_{\beta b}{}^\alpha$ in terms of $F_{abcd}$ as given in Eq.
(\ref{Tf=}).

It is especially interesting to see how the gauge field equations
appear when also the six--superform $A_6$ with the field strength
${\cal F}_7$, Eq.~(\ref{cF7=}), is  introduced \cite{Lechner93}.
Studying (\ref{DF7}), one finds, in addition to $D_\alpha
F_{a_1...a_7}= - 21 i T_{[a_6a_7}{} ^\beta \Gamma_{a_1\ldots
a_5]}{} _{\beta b}{}^\alpha\,$, also the pure bosonic Bianchi
identities
\begin{eqnarray}\label{dF7b}
D_{[c_1} F_{c_2 \ldots c_8]} - {7! \over 4!\, 4!}
F_{[c_1 \ldots c_4} F_{c_5 \ldots c_8]} =0 \;
\end{eqnarray}
and the duality relation for the bosonic fields strength
\begin{eqnarray}\label{F=*F}
F_{c_1 \ldots c_7} = (\ast F_4)_{c_1 \ldots c_7}:= {1\over 4!}
\varepsilon_{c_1 \ldots c_7b_1 \ldots b_4} F^{b_1 \ldots b_4}\; .
\end{eqnarray}
Inserting  the duality relations (\ref{F=*F}) into the bosonic
Bianchi identities for the dual field strength, Eq. (\ref{dF7b}),
 the (superfield generalization of the bosonic) equations of
motion for the three-form gauge field are found,
\begin{eqnarray}\label{d*F}
D_{[c_1} (\ast F_4)_{c_2 \ldots c_8]} - {7! \over 4!\, 4!}
F_{[c_1 \ldots c_4} F_{c_5 \ldots c_8]} =0 \; .
\end{eqnarray}

\subsection{Free differential algebra of  $D=11$ supergravity
}\label{2.3}

A free differential algebra or FDA \cite{Su77,D'A+F,rheoB,Ni83}
(termed Cartan integrable system in \cite{D'A+F}) is an exterior
algebra with constant coefficients generated by a set of forms
that is closed under the action of the exterior differential; the
MC one--forms of a Lie algebra generate the simplest FDA. The
supergravity constraints, Eqs.~(\ref{Ta=})-(\ref{cF7=}), may be
considered as solutions of the equations of a FDA given in terms
of differential forms on superspace. To encode these supergravity
constraints into a FDA one has to (re)define curvatures in such a
way that their definitions include all the terms with derivatives
of forms and the wedge products of forms with constant
coefficients from Eqs.~(\ref{BITa})-(\ref{DF7}) and, instead of
specifying the expressions for these curvatures in terms of
superfields like $F_{abcd}$, $T_{ab}{}^\alpha$,  subject them to
Bianchi identities that are solved by the above supergravity
constraints. Note that the notion of {\it abstract} FDA is more
general than the set of supergravity constraints to which it gives
rise. First, a FDA may be considered as an algebra of forms over
spacetime $M^{11}$ (in this case the FDA curvatures were called
``supersymmetric'' curvatures). But one may also think of it as an
abstract FDA, where all the differential forms characteristic of
$D=11$ supergravity
 \begin{eqnarray}\label{FdAbasis}
E^a \; , \qquad \psi^\alpha \; , \qquad \omega^{ab} \; , \\
\label{FdAbasisA} \qquad A_3   \; ,  \qquad
 A_6 \; , \qquad
\end{eqnarray}
are treated as independent, abstract forms without specifying the
manifold on which they might be defined. For one-forms, this is
tantamount to saying that these forms are defined on a (group)
manifold with a number of coordinates equal to the number of
independent forms, as in the so-called {\it group--manifold} or
{\it rheonomic approach} \cite{Regge,rheoB,D'A+F}.

The FDA of the standard CJS supergravity is defined by the
curvatures of the forms in Eqs.~(\ref{FdAbasis}),
(\ref{FdAbasisA}) \cite{D'A+F}
\begin{eqnarray}\label{CJS:Ta=}
\mathbf{R}^a &:=& DE^a + i \psi^\alpha \wedge \psi^\beta
\Gamma^a_{\alpha\beta} \; ,
\\ \label{CJS:Tf=}
\mathbf{R}^\alpha &:=& T^\alpha=D\psi^\alpha := d\psi^\alpha -
\psi^\beta \wedge \omega_\beta{}^\alpha \; ,
\\ \label{CJS:RL=}
\mathbf{R}^{ab} &:=& R^{ab}=d\omega^{ab} - \omega^{ac}\wedge
\omega_c{}^b \; ,
\\ \label{CJS:R4=}
\mathbf{R}_4 &:=& dA_3 - {1\over 2} \psi^\alpha \wedge \psi^\beta
\wedge \bar{\Gamma}^{(2)}_{\alpha\beta}\; ,
\\ \label{CJS:R7=}
\mathbf{R}_7 &:=& dA_6 + A_3\wedge dA_3 -
 {i\over 2 } \psi^\alpha \wedge \psi^\beta
\wedge \bar{\Gamma}^{(5)}_{\alpha\beta} \; , \qquad
\end{eqnarray}
satisfying the Bianchi identities (\ref{BITa})--(\ref{DF7}), now
written  in terms of $\mathbf{R}^a$, $\dots$, $\mathbf{R}_4$,
$\mathbf{R}_7$,
\begin{eqnarray}\label{BI:Ra}
 {\cal D} \mathbf{R}^a &:=& D \mathbf{R}^a + E^b \wedge  \mathbf{R}_b{}^a
- 2i \psi^\alpha \wedge  \mathbf{R}^\beta \Gamma^a_{\alpha\beta}\equiv 0\; ,
\qquad
\\ \label{BI:Rf}
{\cal D} \mathbf{R}^\alpha  &:=& D \mathbf{R}^\alpha +
{1\over 4} \psi^\beta \wedge  \mathbf{R}^{ab}
\Gamma_{ab}{}_\beta{}^\alpha  \equiv 0 \; , \qquad
\\ \label{BI:RL}
{\cal D} \mathbf{R}^{ab}&:=& DR^{ab} =0 \; , \qquad
\\ \label{BI:F4}
 {\cal D} \mathbf{R}_4 &:=& d\mathbf{R}_4 +
 \psi^\alpha \wedge \mathbf{R}^\beta \wedge
\bar{\Gamma}^{(2)}_{\alpha\beta}
+ {1\over 2} \psi^\alpha \wedge  \psi^\beta \wedge E^b \wedge \mathbf{R}^a
{\Gamma}_{ab}{}_{\alpha\beta} \equiv 0
\; ,
\\ \label{BI:F7}
 {\cal D} \mathbf{R}_7 &:=& d\mathbf{R}_7
- \left(\mathbf{R}_4 + {1\over 2} \psi\wedge
\psi\wedge\bar{\Gamma}^{(2)}\right) \wedge \left(\mathbf{R}_4 +
{1\over 2} \psi\wedge \psi\wedge\bar{\Gamma}^{(2)}\right) -
\nonumber \\ &-& i \psi^\alpha \wedge \mathbf{R}^\beta \wedge
\bar{\Gamma}^{(5)}_{\alpha\beta} +
 {i\over 2\; 4!} \psi^\alpha \wedge  \psi^\beta \wedge E^{c_4} \wedge
 \ldots \wedge  E^{c_1} \wedge  \mathbf{R}^a
{\Gamma}_{ac_1\ldots c_4}{}_{\alpha\beta} + \nonumber \\
&& \qquad + {1\over 4} \psi^\alpha \wedge  \psi^\beta \wedge
\psi^\gamma \wedge  \psi^\delta \wedge
\bar{\Gamma}^{(2)}_{\alpha\beta} \wedge
\bar{\Gamma}^{(2)}_{\gamma\delta} \equiv 0 \; .
\end{eqnarray}

In this abstract FDA framework, the counterpart of the complete
set of the superspace constraints Eqs. (\ref{Ta=})--(\ref{cF7=})
can be written as
\begin{eqnarray}\label{fdTa=}
\mathbf{R}^a &=& 0 \; , \\\label{fdF4=} {\mathbf R}_4 &=& F_4 :=
{1 \over 4! } E^{c_4} \wedge \ldots \wedge E^{c_1} F_{c_1\ldots
c_4} \; ,
\\ \label{fdF7=}  {\mathbf R}_7 &=& F_7 :=
 {1 \over 7! } E^{c_7} \wedge \ldots \wedge E^{c_1}
F_{c_1\ldots c_7}   \; , \qquad
\end{eqnarray}
plus  more complicated expressions for
$\mathbf{R}^\alpha=T^\alpha$ and $\mathbf{R}^{ab}= R^{ab}$,  Eqs.
(\ref{Tf=}), (\ref{RL=}), which can be shortened introducing the
notation
\begin{eqnarray} \label{t1=0}
t_1{}_\beta{}^\alpha &=& E^b t_{b\beta}{}^\alpha :=  {i \over 18}
E^a \left(F_{ac_1c_2c_3}\Gamma^{c_1c_2c_3} + {1\over 8}
F^{c_1c_2c_3c_4} \Gamma_{a c_1c_2c_3c_4}\right)_\beta{}^\alpha \;
,
\end{eqnarray}
in which case they read
\begin{eqnarray}\label{FDA:Tf=}
\mathbf{R}^\alpha & := & T^\alpha=\psi^\beta \wedge
t_1{}_\beta{}^\alpha + {1 \over 2} E^a \wedge E^b T_{ba}{}^\alpha
\; ,
\\ \label{FDA:RL=} \mathbf{R}^{ab}&:=& R^{ab}= 2i \psi^\alpha \wedge
\psi^\beta t^a{}_{(\alpha}{}^\gamma \Gamma_{\beta\gamma}^{b} - E^a
\wedge \psi ^\alpha \left( iT^{ab\beta}\Gamma_{c}{}_{\beta\alpha}
- 2i T_c{}^{[a \, \beta} \Gamma^{b]}{}_{\beta\alpha} \right) +
\qquad
 \nonumber
\\ && \hspace{8cm} + {1 \over 2} E^d \wedge E^c R_{cd}{}^{ab} \; . \qquad
\end{eqnarray}
Eqs. (\ref{fdTa=})--(\ref{fdF7=}), (\ref{FDA:Tf=}),
(\ref{FDA:RL=})
 may be looked at as a solution of the Bianchi identities
(\ref{BI:Ra})-(\ref{BI:F7}). Their pull--back to spacetime
$M^{11}$ or to a bosonic arbitrary eleven-dimensional surface
${\cal M}^{11}\subset\Sigma^{(11|32)}$ in superspace or, even, in
a larger supergroup manifold with more coordinates, can be
obtained from the group--manifold or rheonomic action,
$S=\int_{{\cal M}^{11}}{\cal L}_{11}[E^a, \psi^\alpha,
\omega^{ab}, A_3 , F_{a_1a_2a_3a_4}]$ \cite{D'A+F}, which we
discuss now.

\subsection{First order action for CJS supergravity
with `elementary' $A_3$ field}\label{2.4}

\subsubsection{First order component action}

The first order action for CJS $D=11$ supergravity,
\begin{eqnarray}\label{S11:=}
S=\int_{{M}^{11}}{\cal L}_{11}[E^a, \psi^\alpha, \omega^{ab}, A_3
, F_{a_1a_2a_3a_4}] \quad ,
\end{eqnarray}
is the integral over eleven--dimensional spacetime $M^{11}$ of the
eleven--form ${\cal L}_{11}$  \cite{D'A+F,J+S99}
\begin{eqnarray}\label{L11:=}
{\cal L}_{11} &=& {1\over 4} R^{ab}\wedge E^{\wedge 9}_{ab} -
D\psi^\alpha \wedge \psi^\beta   \wedge
\bar{\Gamma}^{(8)}_{\alpha\beta} +  {1\over 4}  \psi^\alpha \wedge
\psi^\beta   \wedge (T^a + i/2 \, \psi \wedge \psi \, \Gamma^a)
\wedge E_a
\wedge \bar{\Gamma}^{(6)}_{\alpha\beta}  + \nonumber \\
&+&
(dA_3- a_4) \wedge (\ast F_4 + b_7) +  {1\over 2} a_4 \wedge b_7  -
 {1\over 2} F_4 \wedge \ast F_4  -
{1\over 3} A_3 \wedge dA_3\wedge dA_3 \; ,
\end{eqnarray}
where, following \cite{J+S99}, we have denoted
\begin{eqnarray}\label{a4:=}
a_4&:=& {1\over 2} \psi^\alpha \wedge \psi^\beta   \wedge
\bar{\Gamma}^{(2)}_{\alpha\beta}:= - {1\over 4} \psi^\alpha \wedge
\psi^\beta   \wedge E^a \wedge  E^b {\Gamma}_{ab\; \alpha\beta} \;
,  \\ \label{b7:=} b_7&:=&  {i\over 2} \psi^\alpha \wedge
\psi^\beta   \wedge \bar{\Gamma}^{(5)}_{\alpha\beta}:= {i\over 2^.
5!} \psi^\alpha \wedge \psi^\beta   \wedge E^{a_1} \wedge  \ldots
\wedge  E^{a_5} {\Gamma}_{{a_1}  \ldots {a_5} \alpha\beta} \; ,
\qquad
\end{eqnarray}
and introduced {\it purely bosonic} forms $F_4$, $\ast F_4$,
constructed from the auxiliary (zero-form) antisymmetric tensor
$F_{abcd}$ (see also (\ref{fdF4=}), (\ref{fdF7=}))
\begin{eqnarray}\label{F4:=gen}
F_4&:=& {1\over 4!} E^{a_4} \wedge \ldots \wedge E^{a_1} F_{a_1\ldots a_4} \; ,
\\ \label{*F4:=gen}
\ast F_4&:=& - {1\over 4!} E^{\wedge 7}_{a_1\ldots a_4}  F^{a_1\ldots a_4}
 \equiv    {1\over 7!4!} E^{b_7} \wedge \ldots \wedge E^{b_1}
 \; \varepsilon_{b_1\ldots b_7a_1\ldots a_4} F^{a_1\ldots a_4}\; . \qquad
\end{eqnarray}
We also use the compact notation (see Eq.(\ref{Gammaq}))
\begin{eqnarray}
\label{Gammak:=} \bar{\Gamma}^{(k)}_{\alpha\beta} &:=&\!\! {1\over
k!} E^{a_k} \wedge \ldots \wedge E^{a_1} \Gamma_{a_1 \ldots a_k}
{}_{\alpha\beta}:= {(-1)^{k(k-1)/2}\over k!}
\bar{\Gamma}^{(1)}{}_\alpha{}^{\beta_1}\wedge
\bar{\Gamma}^{(1)}{}_{\beta_1}{}^{\beta_2}\wedge  \ldots \wedge
\bar{\Gamma}^{(1)}{}_{\beta_{k-1} \beta} \qquad \nonumber \\ {}
\end{eqnarray}
and
\begin{eqnarray}
\label{E11-n:=} E^{\wedge (11-k)}_{a_1\ldots a_k} := {1\over
(11-k)!} \varepsilon_{a_1\ldots a_kb_1\ldots b_{11-k}}
E^{b_1}\wedge \ldots \wedge E^{b_{11-k}} \; .
\end{eqnarray}
(In the notation of \cite{J+S99},  $E^{\wedge (11-k)}_{a_1\ldots
a_k}:= \Sigma_{a_1\ldots a_k}$ and
$\bar{\Gamma}^{(k)}_{\alpha\beta}:= (-)^{k(k-1)/2}
\gamma^{(k)}_{\alpha\beta}$).

We remark that the Hodge star defined on the purely bosonic
form/tensors does not produce any problem in extending
(\ref{L11:=}) to an eleven--superform on superspace. This allows
for a `rheonomic' treatment of the action (\ref{S11:=}),
(\ref{L11:=}) \cite{Regge,D'A+F,rheoB}. In it, the Lagrangian form
(\ref{L11:=}) is defined on a standard $D=11$ superspace or even
on a larger `supergroup manifold' and ${M}^{11}$ becomes an
arbitrary bosonic surface ${\cal M}^{11}$  in that manifold. See
Sec.~\ref{2.5} for further discussion.

\subsubsection{Equations of motion for $A_3$ and $F_{abcd}$}

Let us denote the eight--form appearing as the variation of the
action (\ref{S11:=}) with respect to $A_3$  by ${\cal G}_8$, {\it
i.e.}
\begin{eqnarray}\label{defcG}
\delta_{A} S= \int {\cal G}_8 \wedge \delta A_3 \; , \qquad
{\delta S \over \delta A_3} := {\cal G}_8 \; .
\end{eqnarray}
{}From Eqs.  (\ref{S11:=}), (\ref{L11:=}) one reads
\begin{eqnarray}\label{defcG=}
{\cal G}_8 &=& d( \ast F_4 +b_7 - A_3\wedge dA_3) \; ,
\end{eqnarray}
and thus the equation of motion for free supergravity in
differential form is
\begin{eqnarray}\label{eqmA3=}
 {\cal G}_8 &=& d( \ast F_4 +b_7 -
A_3\wedge dA_3) =0 \; .
\end{eqnarray}
This includes the auxiliary field  $F_{abcd}= F_{[abcd]}$ (see
(\ref{*F4:=gen})) which on the mass shell is identified with the
covariant field strengths  \cite{D'A+F,J+S99}. Indeed, the variation
of the action with respect to this field has the form
\begin{eqnarray}\label{deFS}
\delta_{F} S &=& \int (dA_3- a_4-F_4) \wedge \ast \delta F_4 =
\\ &=& - {1\over 4!} \int   (dA_3- a_4-F_4) \wedge
E^{\wedge 7}_{a_1\ldots a_4} \, \delta  F^{a_1\ldots a_4}\; . \qquad
\end{eqnarray}
Hence,  the equation of motion $\delta S/\delta  F^{a_1\ldots
a_4}=0$ can be written as
\begin{eqnarray}\label{cdA-=0}
\ast {\delta S \over \delta F_4} =  (dA_3- a_4-F_4)
 =0 \; .
\end{eqnarray}
Notice that Eq. (\ref{cdA-=0}) (see Eq. (\ref{a4:=})) formally
coincides with the FDA relations (\ref{CJS:R4=}) after the
solution of Bianchi identities (\ref{fdF4=}) is used.

\subsubsection{Other equations of motion}

The variation of the action (\ref{S11:=}), (\ref{L11:=}) with
respect to the spin connection gives
\begin{eqnarray}\label{varS=Ta}
&& {\delta {S}_{11}\over \delta{\omega}^{ab}} = {1\over 4}
E^{\wedge 8}_{abc} \wedge (T^c + i\psi^{\alpha}\wedge
\psi^{\beta}\, \Gamma^c_{\alpha\beta}) = 0 \; \quad \Rightarrow
\quad T^a := DE^a = - i\psi^{\alpha}\wedge \psi^{\beta}\,
\Gamma^a_{\alpha\beta} \; . \qquad
\end{eqnarray}
This clearly gives the pull--back of the FDA relation
(\ref{CJS:Ta=}) with (\ref{fdTa=}) for the forms defined on
$M^{11}$ (or defined on a larger superspace but  pulled back on
${M}^{11}$). Taking in mind the algebraic equations
(\ref{varS=Ta}), (\ref{cdA-=0}), one finds that the fermionic
equation for CJS supergravity may be written in the compact form
\cite{J+S99}
\begin{eqnarray}\label{varS=RS}
{\delta {S}_{11}\over \delta{\psi}^{\alpha}} =0 \; \quad
\Rightarrow \qquad  \Psi_{10\; \beta}  &:=& \hat{{\cal D}}
\psi^\alpha \wedge \bar{\Gamma}^{(8)}_{\alpha\beta}  =0 \; ,
\qquad
\end{eqnarray}
in terms of a generalized holonomy connection \cite{Duff03,Hull03}
(see Eq. (\ref{t1=0}) above)
\begin{eqnarray}\label{ghDpsi=}
\hat{{\cal D}} \psi^\alpha &:=& d\psi^\alpha - \psi^\beta \wedge
w_\beta{}^\alpha \equiv d\psi^\alpha - \psi^\beta \wedge
(\omega_\beta{}^\alpha + t_1{}_\beta{}^\alpha) \; ,
\end{eqnarray}
\begin{eqnarray}\label{w=om+t1}
w_\beta{}^\alpha &:=& \omega_\beta{}^\alpha + t_1{}_\beta{}^\alpha \; ,
\quad  \omega_\beta{}^\alpha = {1\over 4} \omega^{ab}\Gamma_{ab}{}_\beta{}^\alpha \; ,
\\ \label{t1=gh}
t_1{}_\beta{}^\alpha &=& {i\over 18} E^a \left({{}\over {}}
F_{ab_1b_2b_3} \Gamma^{b_1b_2b_3}{}_\beta{}^\alpha + {1\over 8}
F^{b_1b_2b_3b_4} \Gamma_{ab_1b_2b_3b_4}{}_\beta{}^\alpha \right)
\; .
\end{eqnarray}
 The explicit form of the Einstein equation for $D=11$
supergravity
\begin{eqnarray}
 \label{EqmE=}
M_{10\, a}  &:=&  R^{bc} \wedge E^{\wedge 8}_{abc} + \ldots =0 \;
\;
\end{eqnarray}
will not be needed in this paper. It can be found in \cite{J+S99}
in similar differential form notation.

\subsection{Rheonomic approach and `generalized action principle': \\ a way from first order
component action to superspace supergravity}\label{2.5}

\subsubsection{The rheonomic action for supergravity as
gauge equivalent to the first order component action}

As known already from \cite{D'A+F}, the action (\ref{S11:=}),
(\ref{L11:=}) may give rise to  the so--called rheonomic action
\cite{Regge,rheoB}. This allows, starting  from a component first
order action, to arrive at the set of superspace supergravity
constraints (see also \cite{BAIL01} for a brief selfcontained
discussion).

The {\it rheonomic action} is obtained by replacing in the action
(\ref{S11:=}), (\ref{L11:=}) all the forms on spacetime $E^a(x)$,
$\psi^\alpha(x)$, $A_3(x)$, $\omega^{ab}(x)$
 (including zero--forms or fields $F_{abcd}(x)$) by  superforms
 (superfields) on the standard superspace $\Sigma^{(11|32)}$,
 Eqs. (\ref{S11Ebf}), (\ref{S11EA}), (\ref{S11A3}),
(\ref{S11dfa}), taken on a bosonic eleven--dimensional surface
${\cal M}^{11}$ in  $\Sigma^{(11|32)}$,
\begin{eqnarray}
 \label{cM11}
{\cal M}^{11}\; & {} : {}& Z^M=\tilde{Z}^M (x^\mu) = (x^{\mu} ,
\tilde{\theta}^{\check{\alpha}}(x))\; ,  \qquad
\end{eqnarray}
\begin{eqnarray}
\left\{
\begin{matrix}
E^a(x)
\cr
\psi^\alpha(x)
\cr
A_3(x)
\cr
F_{abcd}(x)
\end{matrix}\right\}
 \quad \mapsto \quad \left\{\begin{matrix} E^a(\tilde{Z}(x)) \cr
\psi^\alpha(\tilde{Z}(x)) \cr A_3(\tilde{Z}(x)) \cr
F_{abcd}(\tilde{Z}(x))
\end{matrix}\right\} \;
 = \; \left\{\begin{matrix} E^a(x, \tilde{\theta}(x)) \cr
 E^\alpha(x,\tilde{\theta}(x))
\cr
A_3(x, \tilde{\theta}(x))
\cr
F_{abcd}(x, \tilde{\theta}(x))
\end{matrix} \right\} \; . \hspace{-0.8cm} \nonumber \\ {} \hspace{-2cm}
\end{eqnarray}
In this way, the rheonomic action of the $D=11$ supergravity (see
\cite{D'A+F,rheoB}) is given by
\begin{eqnarray}
\label{S11cM} S_{11}^{rh} \! & = & \! \int_{{\cal M}^{11}} {\cal
L}_{11} (x^\mu , \theta^{\check{\alpha}}) := \int_{{M}^{11}} {\cal
L}_{11} (x^{\mu} , \tilde{\theta}^{\check{\alpha}}(x)) \, , \qquad
\end{eqnarray}
where ${\cal L}_{11} (Z^M)\equiv {\cal L}_{11} (x^\mu ,
\theta^{\check{\alpha}})$ is given by Eq. (\ref{L11:=}), where all
the forms on spacetime are replaced  by the superforms
(\ref{S11Ebf}), (\ref{S11EA}), (\ref{S11A3}),
 (\ref{S11dfa}) on the
standard superspace, $\; {\cal L}_{11} (\tilde{Z}^M(x))\equiv
{\cal L}_{11} (x^{\mu} , \tilde{\theta}^{\check{\alpha}}(x))$ is
the pull--back of ${\cal L}_{11} (Z^M)$ by the map $x\mapsto
\tilde{Z}(x)$ in (\ref{cM11}) to the spacetime $M^{11}$. Thus, in
essence, {\it a `rheonomic' action is given by the integral of a
differential $D$--form defined on a $D$--dimensional surface
embedded in a larger manifold (here superspace) and where not only
the fields, but also the surface itself, are varied}.

In principle, one could  also consider ${\cal M}^{11}$ embedded
into the superPoincar\'e group manifold, thus including the
Lorentz group coordinates; the number of independent one--forms
and the number of coordinates would then coincide. However, this
would give nothing new in our perspective\footnote{The inclusion
of the Lorentz group coordinates might be, however, relevant in a
different context as {\it e.g.}, in the search for a formulation
of higher dimensional supergravity in Lorentz--harmonic superspace
(see \cite{BZ,bpstv}) in a way similar to the original `internal'
harmonic superspace approach of \cite{GIKOS}.}. For the standard
$D=11$ supergravity a complete correspondence between different
differential forms and coordinates is impossible due to the
independent three--form field $A_3$. Thus, the apparent lack of
gauge theory treatment  of $D=11$ supergravity makes unfeasible to
complete the `rheonomic' or `group manifold' programme of
\cite{Regge} for this case. This becomes possible if, following
\cite{D'A+F}, one expresses $A_3$ in terms of products of
one--forms. We will discuss this in Sec.~5.

Varying this `rheonomic action' (\ref{S11cM}) with respect to
differential forms one obtains a set of equations like Eqs.
(\ref{eqmA3=}), (\ref{cdA-=0}), (\ref{varS=Ta}), (\ref{varS=RS}),
(\ref{EqmE=}), but for forms replaced  by the superforms on the
surface ${\cal M}^{11}$ of  Eq. (\ref{cM11}),
\begin{eqnarray}\label{dAM=}
dA_3(\tilde{Z}(x)) &=& a_4(\tilde{Z}(x))+ F_4(\tilde{Z}(x)) \; , \qquad
\\
\label{TaM=} T^a (\tilde{Z}(x)) &=&  -
iE^{\alpha}(\tilde{Z}(x))\wedge E^{\beta}(\tilde{Z}(x))\,
\Gamma^a_{\alpha\beta} \; , \qquad
\\
\label{cGM=} {\cal G}_8(\tilde{Z}(x))&=&0\; , \quad \\
\label{cRS=} \Psi_{10 \alpha}(\tilde{Z}(x))&=&0\; , \quad M_{10\;
a}(\tilde{Z}(x))= 0\; . \qquad
\end{eqnarray}
Eqs. (\ref{dAM=}), (\ref{TaM=}) are just the expressions of the
supergravity constraints (\ref{cF4=}) and (\ref{Ta=})
 on the surface ${\cal M}^{11}$. In  Eq. (\ref{dAM=}),
 $a_4(\tilde{Z}(x))$ is the four--form $a_4$ (\ref{a4:=})
 on ${\cal M}^{11}$,
\begin{eqnarray}\label{a4M=}
a_4(\tilde{Z}(x))=
{1\over 2} E^\alpha(\tilde{Z}(x)) \wedge
E^\beta(\tilde{Z}(x))   \wedge \bar{\Gamma}^{(2)}_{\alpha\beta}(\tilde{Z}(x))\; . \qquad
\end{eqnarray}

  The action (\ref{S11cM}) also involves the fermionic field
$\tilde{\theta}^{\check{\alpha}}(x)$ specifying the surface ${\cal
M}^{11}\subset \Sigma^{(11|32)}$, Eq. (\ref{cM11}). This field is
treated as a dynamical variable, on the same footing as the
differential (super)forms $E^a$ etc. In the original articles
\cite{Regge,rheoB} (see also \cite{D'A+F}) this corresponds to the
statement that the surface ${\cal M}^{11}$ itself is varied as the
differential form fields are. Thus,
\begin{eqnarray}\label{vM=vtt}
\delta {\cal M}^{11} \leftrightarrow \delta \tilde{\theta}^{\check{\alpha}}(x)\; ,
\end{eqnarray}
and the complete variation of the rheonomic action (\ref{S11cM})
reads
\begin{eqnarray}
\label{vS11cM} \delta S_{11}^{rh} & = & \int_{{\cal M}^{11}}
\delta {\cal L}_{11} (Z) +
 \int_{\delta {\cal M}^{11}} {\cal L}_{11} (Z)
\nonumber \\
&\equiv & \int_{{M}^{11}} \delta {\cal L}_{11}
(Z)\vert_{Z^M=\tilde{Z}^M(x)} + \int_{{M}^{11}} {\partial {\cal
L}_{11}(x, \tilde{\theta}(x)) \over \partial
\tilde{\theta}^{\check{\alpha}}(x)} \;  \delta
\tilde{\theta}^{\check{\alpha}}(x) \; . \qquad
\end{eqnarray}

The complete set of the equations of motion that follow from the
rheonomic action (\ref{S11cM}) includes, in addition to
(\ref{dAM=}), (\ref{TaM=}), (\ref{cGM=}), (\ref{cRS=}),  the
Euler-Lagrange equation for the fermionic field
$\tilde{\theta}^{\check{\alpha}}(x)$. It is given by
\begin{eqnarray}\label{Eqmtth}
{\delta S_{11}^{rh} \over \delta
\tilde{\theta}^{\check{\alpha}}(x)} &:=&
 {\partial {\cal L}_{11}(x, \tilde{\theta}(x)) \over \partial
\tilde{\theta}^{\check{\alpha}}(x)} =0  \; .
\end{eqnarray}
However, this new equation (\ref{Eqmtth}) is satisfied identically
when Eqs. (\ref{dAM=})--(\ref{cRS=}) are taken into account (see
\cite{Regge,rheoB}). To see this  one first notices that the
variation of the  Lagrangian form ${\cal L}_{11}$ can be written
as a Lie derivative ${L}_\delta = di_\delta + i_\delta d$, where
$i_\delta$ is the inner product with respect to the vector field
that determines the variation. It satisfies Leibniz's rule,
\begin{eqnarray}
\label{idOmOm} i_\delta (\Omega_q \wedge \Omega_p ) &=& \Omega_q
\wedge i_\delta \Omega_p + (-)^p (i_\delta \Omega_q) \wedge
\Omega_p \; , \qquad
\end{eqnarray}
 for any $p$($q$)--form $\Omega_{p}$ ($\Omega_{q}$) (as the
 one--forms $E^a$, $\psi^\alpha$, and the three--form $A_3$).
The variation of the rheonomic action (\ref{vS11cM}) is given by the
pull--back of
\begin{eqnarray}\label{vL=LL}
\delta {\cal L}_{11} (Z) = i_\delta d{\cal L}_{11} (Z) + d(i_\delta {\cal L}_{11} (Z))
\end{eqnarray}
where the second term may be ignored in $\delta S_{11}^{rh}$ when
the surface ${\cal M}^{11}$ has no boundary, $\partial{\cal
M}^{11}=\emptyset$ [notice that this is not the case for
Ho\v rava--Witten heterotic M-theory \cite{HW96}, but we do  not
consider this case here]. Thus, $\delta S_{11}^{rh}  = \int_{{\cal
M}^{11}} i_\delta d{\cal L}_{11} (Z) + \int_{{M}^{11}}
\partial {\cal L}_{11}(x, \tilde{\theta}(x)) /
\partial \tilde{\theta}^{\check{\alpha}}(x)\; \delta
\tilde{\theta}^{\check{\alpha}}(x)$ and all the equations of
motion (\ref{dAM=}), (\ref{TaM=}), (\ref{cGM=}), (\ref{cRS=})
follow from
\begin{eqnarray}\label{Eqm}
i_\delta d{\cal L}_{11} (Z)\vert_{{\cal M}^{11}} =0  \; .
\end{eqnarray}
Reciprocally, Eq. (\ref{Eqm}) is satisfied {\it for any variation
$\delta$} if the equations of motion (\ref{dAM=}), (\ref{TaM=}),
(\ref{cGM=}), (\ref{cRS=}) are taken into account. Now, the second
term in (\ref{vS11cM}) comes from a particular fermionic general
coordinate transformation of the superform, ${\cal L}_{11}(x,
{\theta} + \delta {\theta})- {\cal L}_{11}(x, {\theta})$ on ${\cal
M}^{11}$, and, hence, is also given by the Lie derivative, ${\cal
L}_{11}(x, {\theta} + \delta {\theta})- {\cal L}_{11}(x,
{\theta})= - (i_\delta d + di_\delta){\cal L}_{11}(x, {\theta})$,
but now with respect to the vector field defining  the variation
$\delta \tilde{\theta}^{\check{\alpha}}(x)$ of the fermionic field
$\tilde{\theta}^{\check{\alpha}}(x)$,
\begin{eqnarray}\label{vfL=LfL}
\int {\partial  {\cal L}_{11}(x, \tilde{\theta}(x)) \over \partial
\tilde{\theta}^{\check{\alpha}}(x)}
\delta\tilde{\theta}^{\check{\alpha}}(x) = \int i_{\delta
\tilde{\theta}^{\check{\alpha}}(x)}  d{\cal L}_{11}(x,
\tilde{\theta}(x))\; .
\end{eqnarray}
As a result, Eq. (\ref{Eqmtth}) reads ({\it cf.} Eq. (\ref{Eqm}))
\begin{eqnarray}\label{Eqmtth1}
 { \delta S^{rh}_{11} \over \delta \tilde{\theta}^{\check{\alpha}}(x)}
 \delta\tilde{\theta}^{\check{\alpha}}(x) =
 i_{\delta \tilde{\theta}^{\check{\alpha}}(x)}
 d {\cal L}_{11}(x, \tilde{\theta}(x))=0  \;
\end{eqnarray}
so that it  is satisfied identically due to (\ref{Eqm}) {\it
i.e.}, when the equations of motion (\ref{dAM=})--(\ref{cRS=}) are
taken into account. This dependence of the equation of motion
(\ref{Eqmtth1}) is just the Noether identity reflecting the
existence of a gauge symmetry that acts additively on the
fermionic function $ \tilde{\theta}^{\check{\alpha}}(x)$, $\delta
\tilde{\theta}^{\check{\alpha}}(x)= \beta^{\check{\alpha}}(x)$.
This fermionic gauge symmetry is the symmetry under arbitrary
`deformations' (changes) of
 the bosonic surface
${\cal M}^{11}$ in superspace,
\begin{eqnarray}\label{Msymm}
\delta \tilde{\theta}^{\check{\alpha}}(x)= \beta^{\check{\alpha}}(x) \; \Leftrightarrow \qquad
\delta {\cal M}^{11}= arbitrary \; .
 \end{eqnarray}
This independence on the location of the bosonic surface ${\cal
M}^{11}$ in $\Sigma^{(11|32)}$ is the basis for the formulation of
the rheonomic or {\it ``generalized action'' principle}.

Let us stress that, although the above proof uses on--shell
arguments (usual in the language of the second Noether theorem),
as Eq. (\ref{Eqm}) collects the equations of motion
(\ref{dAM=})--(\ref{cRS=}), the variation (\ref{Msymm}) provides a
true gauge symmetry of the action (\ref{S11cM})\footnote{What
happens is that the variation of the
$\tilde{\theta}^{\check{\alpha}}(x)$, which enters as a parameter
of superforms,  is compensated by the appropriate variations of
the functions ($E_M^a$ {\it etc.}) in these superforms; see
\cite{BAIL01} for further discussion.}. In particular the
transformations of this symmetry can be used to set
$\tilde{\theta}^{\check{\alpha}}(x)=0 $ {\it i.e.}, to identify
${\cal M}^{11}$ with the bosonic body $M^{11}$ of superspace.
Hence  one sees that the rheonomic action of supergravity is gauge
equivalent to the component first order action defined in terms of
the spacetime component fields.

\subsubsection{Generalized action principle: rheonomic action plus
 `lifting' to superspace}

What is new in the rheonomic action with respect  to the component
one is that it produces  equations that are valid on an arbitrary
surface ${\cal M}^{11}$ in  standard superspace. Moreover, it has
a gauge symmetry that allows for arbitrary changes of this
surface, Eq. (\ref{Msymm}). As a result the (differential
(super)form) equations of motion, Eqs. (\ref{dAM=})--(\ref{cRS=}),
are valid on an {\it arbitrary} surface in  superspace.
Furthermore, since the set of all these surfaces span the whole
superspace, this suggests that one may try to extend or `lift'
these equations from a surface ${\cal M}^{11}$ in superspace  to
superspace itself, {\it i.e.} to substitute the fermionic
superspace coordinates $\theta$ and $d\theta$ for the fermionic
coordinate functions $\tilde{\theta}(x)$ and $d\tilde{\theta}(x)$
in them. This procedure, the so--called rheonomic {\it lifting},
is not a consequence of the rheonomic action (\ref{S11cM}), but
rather an additional step, the consistency of which needs to be
checked. Although such a consistency is not guaranteed (see
\cite{BAIL01} and refs. therein for a discussion), it works for
the standard CJS supergravity:  lifting Eqs. (\ref{dAM=}),
(\ref{TaM=}) to the standard superspace one arrives at the
standard superspace $\Sigma^{(11|32)}$  constraints, Eqs.
(\ref{cF4=}) and (\ref{Ta=}). All other constraints (\ref{Tf=}),
(\ref{RL=}) as well as the consequences of the superfield
relations that follows from the lifting of the dynamical equations
(\ref{cGM=}), (\ref{cRS=}) can be reproduced by checking the
consistency of the superspace constraints (\ref{Ta=}) and
(\ref{cF4=}), {\it i.e.} by studying the Bianchi identities (see
Sec~\ref{sugraconstr}).

Thus, and precisely  in this sense, the rheonomic action
(\ref{S11cM}) provides a bridge between the spacetime component
action and the standard superspace formulation of supergravity.
Considered together with the second step of rheonomic lifting this
action leads to the `generalized action principle' \cite{Regge}
(see also  \cite{bsv}) which reproduces the $D=11$ superspace
supergravity constraints. In Sec. 5 we will see that the rheonomic
action with a composite $A_3$ (Sec. 4) leads us naturally to an
enlarged superspace with additional bosonic and fermionic
coordinates in the sense that ${\cal M}^{11}$ may be understood as
an arbitrary surface in this superspace. But before turning to the
composite structure of $A_3$, let us discuss the relation of the
FDA in Sec. 2.3 with the rigid standard superspace, the supergroup
manifold of the supertranslation group, and show that the $A_3$
(super)form is the potential three--form of a nontrivial
Chevalley--Eilenberg four--cocycle on the standard supersymmetry
algebra $\mathfrak{E}^{(11|32)}$.

\setcounter{equation}0
\section{Rigid  superspaces as supergroup manifolds:
{}From FDA to Lie algebras, CE cocycles and enlarged
 superspaces}

To make clear the relation of a FDA with a supergroup manifold,
let us consider the one defined by (\ref{CJS:Ta=}),
(\ref{CJS:Tf=}), (\ref{CJS:RL=}) setting the curvatures equal to
zero, $\mathbf{R}^a = 0$, $\mathbf{R}^\alpha=0$,
$\mathbf{R}^{ab}=0$. The resulting equations are  the
Maurer--Cartan (MC) equations of the superPoincar\'e algebra
\begin{eqnarray}\label{sP:Ta=}
dE^a &=& E^b \wedge \omega_b{}^a
- i \psi^\alpha \wedge \psi^\beta \Gamma^a_{\alpha\beta} \; ,
\\ \label{sP:Tf=}
d\psi^\alpha &=& \psi^\beta \wedge \omega_\beta{}^\alpha  \; ,
\\ \label{sP:RL=}
d\omega^{ab} &=& \omega^{ac}\wedge \omega_c{}^b \; .
\end{eqnarray}
One may  easily solve these equations by
\begin{eqnarray}\label{sP:Pi}
\omega^{ab}=0 \; , \qquad \psi^\alpha= \Pi^\alpha =
d\theta^{\check{\alpha}} \delta_{\check{\alpha}}{}^{{\alpha}}:=
d\theta^{{\alpha}}\; , \qquad
\nonumber \\
E^a= \Pi^a := dx^\mu \delta_\mu{}^a - i d\theta \Gamma^a \theta \;
; \qquad
\end{eqnarray}
where $\Pi^a$, $\Pi^\alpha$ are the (MC) one--forms of
 the standard supersymmetry algebra $\mathfrak{E}= \mathfrak{E}^{(11|32)}$,
 \begin{equation}
\{ Q_\alpha, Q_\beta \} = \Gamma^a_ {\alpha\beta} P_a \;.
\end{equation}
 Considered as forms on
rigid superspace ($\Sigma^{(D|n)}$ in general), one identifies $x^a$
and $\theta^\alpha$ with the coordinates $Z^M=(x^a,\theta^\alpha)$
of this superspace. Notice that the standard $D$=11 {\it rigid}
superspace is the group manifold of the supertranslations group
$\Sigma^{(11|32)}$. When $E^a=\Pi^a$ and $\psi^\alpha=\Pi^{\alpha}$
are forms on spacetime, $x^a$ are still spacetime coordinates while
$\theta^\alpha$ are Grassmann functions,
$\theta^\alpha=\theta^\alpha(x)$, the Volkov-Akulov Goldstone
fermions \cite{VA72}.

Thus, when the curvatures in Eqs.  (\ref{CJS:Ta=}),
(\ref{CJS:Tf=}), (\ref{CJS:RL=}) are set to zero, the one--form
gauge fields $E^a$ and $\psi^\alpha$ of supergravity become
identified with the MC forms of $\mathfrak{E}^{(11|32)}$ which
have a natural representation as one--forms on rigid  superspace
$\Sigma^{(11|32)}$. If we now look at the $D=11$ supergravity FDA,
it is also natural to ask what is the `flat' limit of the
three--form gauge field $A_3$. To answer this question we set
further $\mathbf{R}_4 =0$ in (\ref{CJS:R4=}) and find $dA_3= w_4$
where $w_4$ is the `flat value' of the bifermionic form $a_4$ in
Eq. (\ref{a4:=}),
\begin{eqnarray} \label{R4=0}
dA_3 = w_4:= - {1\over 4} \Pi^\alpha \wedge \Pi^\beta \wedge \Pi^a
\wedge \Pi^b\; {\Gamma}_{ab \; \alpha\beta}\; .
\end{eqnarray}
 The {\it r.h.s.} of this
equation, $w_4$, is a supersymmetry invariant closed four--form. It
is also exact (trivial) in the de Rahm cohomology, $w_4=
dw_3(x,\theta)$, but the three--form $w_3(x,\theta)$ is not
invariant under rigid supersymmetry transformations \footnote{Note
that, under {\it local} supersymmetry, $\delta_\varepsilon A_3=-
{1\over 2} d\theta^\alpha \wedge \Pi^a \wedge \Pi^b\; {\Gamma}_{ab
\; \alpha\beta}\, \varepsilon^\beta (x) + d\alpha_2\,$, where we
allow for the presence of an arbitrary 2--form $\alpha_2$ which
could be identified with the parameter of the three-form gauge
transformations; for the general FDA case, where the curvatures are
nonvanishing, defining  $\delta_\varepsilon A_3=- {1\over 2}
\psi^\alpha \wedge E^a \wedge E^b\; {\Gamma}_{ab \; \alpha\beta}\,
\varepsilon^\beta (x) + d\alpha_2\,$ one finds that ${\mathbf{R}}_4$
is invariant under local supersymmetry.} since $w_3(x,\theta)$
involves the Grassmann coordinates $\theta$ explicitly (not through
the MC one--forms $\Pi^a$ and $\Pi^\alpha$). In fact, the superspace
three--form $w_3(x,\theta)$ is well known, as its pull--back to the
worldvolume ${\cal W}^3$ is the Wess--Zumino term in the $D$=11
supermembrane action \cite{BST}.
 Thus, $w_4:= dw_3$ ($dA_3$) defines as a nontrivial
$\mathfrak{E}^{(11|32)}$ CE four--cocycle, since it is not the
exterior derivative of an {\it invariant} form on
$\Sigma^{(11|32)}$. To trivialize this CE cocycle {\it i.e.}, to
write $w_3$ in terms of the MC forms of some superalgebra,
 one may, following the point of
view of \cite{BESE,JdA00}, enlarge the superspace group manifold
$\Sigma$ (superalgebra ${\mathfrak{E}}$) to $\tilde{\Sigma}$
($\tilde{\mathfrak{E}}$) by adding a number of additional bosonic
and fermionic coordinates (generators) (see \cite{Lett} and Sec. 6
for further discussion). We will show below that such extended
superspaces [supergroups] $\tilde{\Sigma}$ exist, with 528 bosonic
and 64 fermionic dimensions,  and that they can be identified with
the nontrivial deformations $\tilde{\Sigma}^{(528|32+32)}(s\not=0)$
of the supergroup manifold $\tilde{\Sigma}^{(528|32+32)}(0)$
\cite{Lett}, the algebra of the latter being related to an {\it
expansion} \cite{JdA02} of $osp(1|32)$.

Before  turning to the trivialization of the ${\Sigma}^{(11|32)}$
four--cocycle on the Lie superalgebra $\tilde {\mathfrak
E}(s)=\tilde {\mathfrak E}^{(528|32+32)}(s)$ and to the  equivalent
problem of finding the composite structure of the  $A_3$ field of CJS
supergravity in terms of $\tilde {\mathfrak E}(s)$ gauge fields, let
us comment on the `flat' limit of the dual six--form field.
 Setting $\mathbf{R}_7 =0$, one finds from (\ref{CJS:R7=})
\begin{eqnarray}  \label{R7=0}
dA_6 &=&-  A_3\wedge dA_3 + {i\over 2 \cdot 5!} \Pi^\alpha \wedge
\Pi^\beta \wedge \Pi^{a_5} \wedge \ldots \wedge \Pi^{a_1} \,
{\Gamma}_{a_1\ldots a_5\; \alpha\beta} := -A_3 \wedge dA_3 + b_7^0
 \; , \qquad
\end{eqnarray}
where $b^0_7$ is the `flat' value of the bifermionic seven  form
$b_7$, Eq. (\ref{b7:=}).
 The consistency
condition $ddA_6=0$ is satisfied due to (\ref{R4=0}) and the
$D$=11 identity
\begin{equation}
\Gamma_{a \ (\alpha \beta} \Gamma^{ab_1 \cdots b_4}{}_{\gamma
\delta)} = 3 \Gamma^{[b_1 b_2}{}_{(\alpha \beta} \Gamma^{b_3
b_4]}{}_{\gamma \delta)} \; .
\end{equation}
However, $dA_6$ is not a CE seven-cocycle on
$\mathfrak{E}^{(11|32)}$ because, as stated, the three--form
$A_3=w_3(x,\theta)$ is not invariant under the standard
supersymmetry group transformations. This, however, is the case on
the $\tilde{\Sigma}^{(528|32+32)}(s\not=0)$ superspace, where, as
we show in the next section, $dA_3$  itself is `trivialized' {\it
i.e.} $A_3$ is expressed  as a product of the MC one-forms
invariant under the enlarged supersymmetry group
$\tilde{\Sigma}^{(528|32+32)}(s\not=0)$. It would be interesting
to see  whether the cocycle (\ref{R7=0}) is already trivial on the
$\tilde{\Sigma}^{(528|32+32)}(s\not=0)$ superspace or whether a
further extension is needed. Another, equivalent, formulation  of
the same problem is whether the six--form $A_6$ may be expressed,
as $A_3$, in terms of the $\tilde{\Sigma}^{(528|32+32)}(s\not=0)$
gauge fields. This corresponds to looking for an underlying gauge
group structure for the selfdual formulation of $D=11$
supergravity \cite{BBS97} (see also \cite{Lechner93},
\cite{Dualdual2}).

\setcounter{equation}0
\section{Trivialization of the four-cocycle and the underlying
gauge group structure of D=11 supergravity}\label{Sectriv}

The general FDA defined by the set of Eqs.~(\ref{CJS:Ta=}),
(\ref{CJS:Tf=}), (\ref{CJS:RL=}) may be treated as a `gauging' of
the super--Poincar\'e group described by the superPoincar\'e algebra
MC equations (\ref{sP:Ta=}), (\ref{sP:Tf=}), (\ref{sP:RL=}). In
$D=4$, where the supergravity multiplet consists of the graviton and
the gravitino, this allows one to state that supergravity is a gauge
theory of the superPoincar\'e group (see references in \cite{CFGPN}
and \cite{Ha-Tr-Za-04}). The one-forms $E^a(x)$, $\psi^\alpha(x)$,
$\omega^{ab}(x)$ are then treated as gauge fields for local
translations (or general coordinate transformations), local
supersymmetry and Lorentz rotations. However, for $D=11$ CJS
supergravity this is prevented by the presence of an `elementary'
three-form gauge field $A_3(x)$.

\subsection{
The $\tilde {\mathfrak E}^{(528|32+32)}(s)$ family of
superalgebras}

As stated in  \cite{D'A+F}, the problem is whether the FDA
(\ref{CJS:Ta=}), (\ref{CJS:Tf=}), (\ref{CJS:RL=}) may be completed
with a number of additional {\it one}--forms and their curvatures
in such a way that the three-form $A_3$ obeying (\ref{CJS:R4=}) is
constructed from one-forms, becoming composite rather than
fundamental or `elementary'. This problem is equivalent to
trivializing  the $\Sigma^{(11|32)}$ four-cocycle $dA_3=w_4$, Eq.
(\ref{R4=0}), on the algebra of an {\it enlarged} superspace
group.

Indeed, $A_3$ may be constructed in terms of the graviton,
gravitino, an antisymmetric second rank tensor one--form
$B_1^{ab}$, a fifth rank antisymmetric tensor one--form
$B_1^{a_1\ldots a_5}$ and  an additional {\it fermionic spinor}
one--form $\eta_1^\alpha$,
\begin{eqnarray}
\label{A3=A3def2} A_3 = A_3(E^a ,\, \psi^\alpha \; ; \; B_1^{ab},
\, B_1^{abcde},\, \eta_{1\, \alpha} )\;.
\end{eqnarray}
The curvatures of the new forms are defined by
\begin{eqnarray}
\label{cBab} {\cal B}_2^{ab} &=& DB_1^{ab} + \psi^\alpha  \wedge
\psi^\beta \, \Gamma^{ab}_{\alpha\beta} \; , \qquad
\\
\label{cB15} {\cal B}_2^{a_1\ldots a_5} &=& DB_1^{a_1\ldots a_5} +
i \psi^\alpha  \wedge \psi^\beta \,
\Gamma^{a_1\ldots a_5}_{\alpha\beta} \; , \qquad \\
 \label{cDeta}
{\cal B}_{2\alpha } &=& D\eta_{1\alpha} - i \, \delta \, E^a
\wedge \psi^\beta \Gamma_{a\, \alpha\beta} -  \gamma_1 \, B_1^{ab}
\wedge \psi^\beta \Gamma_{ab\, \alpha\beta} -  i \, \gamma_2 \,
B_1^{a_1\ldots a_5} \wedge \psi^\beta \Gamma_{a_1\ldots
a_5\alpha\beta} \; ,  \qquad
\end{eqnarray}
where $D$ is the Lorentz covariant derivative and $\delta$,
$\gamma_1$, $\gamma_2$ are constants to be fixed.

Let us discuss enlarging the FDA of Eqs. (\ref{CJS:Ta=}),
(\ref{CJS:Tf=}) and (\ref{CJS:RL=}) by the one--forms and
curvatures of Eqs. (\ref{cBab}), (\ref{cB15}) and (\ref{cDeta}) in
more detail. The constants in these expressions must obey one
single relation (\ref{idg0}) \cite{D'A+F},
\begin{eqnarray}
\label{id1} \delta + 10 \gamma_1 - 720 \gamma_2= 0 \; ,
\end{eqnarray}
 which comes from the selfconsistency (closure under the exterior
derivative) of eq. (\ref{cDeta}) after using the rest of the FDA
equations, Eqs. (\ref{CJS:Ta=}), (\ref{CJS:Tf=}), (\ref{CJS:RL=}),
(\ref{cBab}), (\ref{cB15}) and the identity
\begin{eqnarray}
 \label{G1G1=G2G2}
\Gamma_{b(\alpha \beta}\Gamma^b_{\gamma\delta )} &=& - {1\over 10}
\Gamma_{ab(\alpha \beta}\Gamma^{ab}{}_{\gamma\delta )} = - {1\over
720} \Gamma_{a_1\ldots a_5(\alpha \beta} \Gamma^{a_1\ldots
a_5}{}_{\gamma\delta )} \; .
\end{eqnarray}

Setting the curvatures in Eqs. (\ref{cBab})--(\ref{cDeta}) equal
to zero and omitting the trivial Lorentz connection, the resulting
equations plus Eqs. (\ref{sP:Ta=})--(\ref{sP:RL=}) become the MC
equations of an enlarged  supersymmetry algebra with the following
nonvanishing (anti)commutators
\begin{eqnarray}\label{QQ}
 \{ Q_\alpha , Q_\beta \} &=& \Gamma^{a}_{\alpha\beta}  P_a + i \Gamma^{ab}_{\alpha\beta} Z_{ab}
+ \Gamma^{a_1\ldots a_5}_{\alpha\beta} Z_{a_1\ldots a_5} \; , \qquad
\\
\label{PQ}
 && [ P_a , Q_\alpha ] = \delta \, \Gamma_{a \;
\alpha\beta}  \, Q^{\prime\beta} \; , \quad
\nonumber \\
 && [ Z_{ab}  , Q_\alpha ] =
 i \gamma_1 \, \Gamma_{ab \;
\alpha\beta} \, Q^{\prime\beta } \; , \quad
\nonumber \\
 && [ Z_{a_1\ldots a_5} ,  Q_\alpha ] = \gamma_2 \,
\Gamma_{a_1\ldots a_5 \; \alpha\beta}  Q^{\prime\beta} \; ,
\\ \label{Q1all=0}
&& \qquad {} \qquad [ Q^{\prime\alpha}\, , \, all \, \}  = 0\; .
\end{eqnarray}
Clearly, as long as the constant $\gamma_1$ (or $\delta$) is
nonvanishing, it can be included in the normalization of the
additional fermionic central generator $Q^{\prime\alpha}$ or,
equivalently, in the one--form $\eta_{1\, \alpha}$ in
(\ref{cDeta}).

Upon solving condition (\ref{id1}) on the constants $\delta$,
$\gamma_1$, $\gamma_2$ in terms of one parameter $s$ (\ref{s-def})
and $\gamma_1$, one writes the algebra (\ref{QQ}), (\ref{PQ}) in
the form of Eq. (\ref{QQ}) and
\begin{eqnarray}\label{Sigma(s)}
[ P_a , Q_\alpha ] &=& 2\gamma_1(s+1) \;  \Gamma_{a \;
\alpha\beta} Q^{\prime\beta} \; , \quad \nonumber
\\ {} [ Z_{a_1a_2} , Q_\alpha ] &=&i\gamma_1
\Gamma_{a_1a_2 \; \alpha\beta} Q^{\prime\beta} \; , \nonumber \\
\quad [ Z_{a_1 \ldots a_5} , Q_\alpha ]&=& 2\gamma_1({s \over 6!}
+ {1 \over 5!}) \Gamma_{a_1 \ldots a_5 \; \alpha\beta}
Q^{\prime\beta} \; .
\end{eqnarray}
Thus, one concludes that the family of the fermionic central
extensions ($[Q^{\prime\alpha}, \, all  \} =0$) of the M-theory
superalgebra described by Eqs. (\ref{QQ}), (\ref{PQ}) is
effectively one--parametric. Following  \cite{Lett} we denote the
family of superalgebras given by Eqs.  (\ref{QQ}) and
(\ref{Sigma(s)}) by $\tilde {\mathfrak E}(s)=\tilde {\mathfrak
E}^{(528|32+32)}(s)$, and by $\tilde {\Sigma}(s)=\tilde
{\Sigma}^{(528|32+32)}(s)$ the associated extended superspace
group manifolds. The MC equations of $\tilde{\mathfrak
E}^{(528|32+32)}(s)$ are given by the above FDA equations for zero
curvatures {\it i.e.},
\begin{eqnarray}\label{MCTa=}
d\Pi^a &=& - i\Pi^{\alpha} \wedge \Pi^{\beta}
\Gamma^a_{\alpha\beta} \; , \qquad \nonumber \\ d\Pi^\alpha &=& 0
\; ,
\\
\label{dB2=} d\Pi^{a_1a_2} &=& - \Pi^\alpha \wedge \Pi^\beta \,
\Gamma^{a_1a_2}_{\alpha\beta} \; , \qquad \\
\label{dB5=}  d\Pi^{a_1\ldots a_5} &=& - i \Pi^\alpha \wedge
\Pi^\beta \, \Gamma^{a_1\ldots a_5}_{\alpha\beta} \; ,  \\
 \label{deta=}
d\Pi^{\prime}_{\alpha} &=& -2\gamma_1 \Pi^\beta \wedge \left(i \,
(s+1) \, \Pi^a \Gamma_{a} + {1\over 2}  \Pi^{ab} \Gamma_{ab} +
  i  \left({s \over 6!} + {1 \over 5!}\right) \,  \,
\Pi^{a_1\ldots a_5} \Gamma_{a_1\ldots a_5} \right)_{\beta\alpha}
. \qquad
\end{eqnarray}

\subsection{The
$\tilde {\mathfrak E}^{(528|32+32)}(0)$ superalgebra and its
associated FDA}

The $D$=11 Fierz identity
\begin{eqnarray}
 \label{II=GG}
\delta_{(\alpha}{}^{\gamma} \delta_{\beta)}{}^{\delta} &=& {1\over
32} \Gamma^a_{\alpha\beta} \Gamma_a^{\gamma\delta} - {1\over 64}
\Gamma^{a_1a_2}{}_{\alpha\beta}\Gamma_{a_1a_2}{}^{\gamma\delta} +
{1\over 32 \cdot 5!} \Gamma^{a_1\ldots a_5}{}_{\alpha\beta}
\Gamma_{a_1\ldots a_5}{}^{\gamma \delta}\;
\end{eqnarray}
allows one to collect the set of one--forms $E^a$, $B^{ab}_1$,
$B^{abcde}_1$ into one symmetric spin--tensor one--form ${\cal
E}^{\alpha\beta}$,
\begin{eqnarray}
 \label{cEff=def}
{\cal E}^{\alpha\beta}&=& {1\over 32}\left(E^a
\Gamma_{a}^{\alpha\beta} - {i\over 2}
B_1^{a_1a_2}\Gamma_{a_1a_2}{}^{\alpha\beta}  + {1\over 5!}
B_1^{a_1\ldots a_5} \Gamma_{a_1\ldots a_5}{}^{\alpha\beta}
\right)\; .
\end{eqnarray}
The curvatures
\begin{eqnarray}
\label{cEab} {\cal R}^{\alpha\beta} &=& D{\cal E}^{\alpha\beta} +
i \psi^\alpha  \wedge \psi^\beta \; , \qquad
\end{eqnarray}
(\ref{CJS:Tf=}), (\ref{CJS:RL=}) of the set of one--forms ${\cal
E}^{\alpha\beta}$, ${\psi}^{\alpha}$, $\omega^{ab}$  satisfy the
Bianchi identities
\begin{eqnarray}\label{BI:Rff}
{\cal D} {\cal R}^{\alpha\beta} &:=& D{\cal R}^{\alpha\beta} + 2
{\cal E}^{\gamma (\alpha } \wedge  {\mathbf R}_{\gamma}{}^{\beta
)}  - 2 i \psi^{(\beta }\wedge \mathbf{R}^{\alpha )} \equiv 0 \; ,
 \nonumber \\ && \quad {\mathbf R}_{\gamma}{}^{\beta } \equiv {1\over
4} {\mathbf R}^{ab} \Gamma_{ab\, \gamma}{}^{\beta } \; ,
\end{eqnarray}
(\ref{BI:Rf}) and (\ref{BI:RL}). This FDA includes the
Lorentz-spin connection $\omega^{ab}$ and its curvature
$\mathbf{R}^{ab}$, Eq. (\ref{CJS:RL=}).

If we move to the flat limit where all curvatures are zero, and set
to zero the spin connection, $\omega^{ab}=0$, the one-forms
$E^a=\Pi^a$, $\psi^\alpha=\Pi^\alpha$ $B_1^{ab}=\Pi^{ab}$,
$B_1^{a_1\ldots a_5}=\Pi^{a_1 \ldots a_5}$ obey the MC equations
(\ref{MCTa=})--(\ref{dB5=}). These can be collected in the compact
expression
\begin{eqnarray}
\label{cEab0} d{\Pi}^{\alpha\beta} =-  i \Pi^\alpha \wedge
\Pi^\beta \;  \quad , \quad d\Pi^\alpha=0  \quad
\end{eqnarray}
clearly exhibiting
 a $GL(n)$ symmetry ($\Pi^{\alpha \beta}$ is the flat limit of
 ${\cal E}^{\alpha\beta}$ in (\ref{cEff=def})).
 This is the automorphism
symmetry of the M-theory superalgebra $\tilde {\mathfrak
E}^{(528|32)}$ which is defined by the MC equations (\ref{cEab0})
and can be obtained from any of the $\tilde {\mathfrak
E}^{(528|32+32)}(s)$ superalgebras just by setting the fermionic
central charge $Q^{\prime\alpha}$  equal to zero. Clearly, none of
the $\tilde {\mathfrak E}^{(528|32+32)}(s)$ superalgebras possess
the full $GL(n)$ automorphism symmetry; for $s\neq 0$ they only
possess the Lorentz one $SO(1,10)$. This automorphism group is
enhanced to $Sp(32)$ when $s=0$ {\it i.e.}, for the special values
of the constants $\delta$, $\gamma_1$, $\gamma_2$ given by
\begin{eqnarray}
\label{sol1} \delta = 2\gamma_1 \quad ,\quad \gamma_2 =
\frac{2}{5!} \gamma_1
 \; \qquad \Leftrightarrow \quad s=0 \; .
\end{eqnarray}
Indeed, the $\tilde {\mathfrak E}^{(528|32+32)}(0)$ algebra MC
equations, Eqs. (\ref{MCTa=})--(\ref{deta=}), can be collected in
\begin{eqnarray}
\label{MC:s=0}  \tilde {\mathfrak E}^{(528|32+32)}(0) \quad :
\qquad \begin{cases} d{\Pi}^{\alpha\beta} = -  i \Pi^\alpha \wedge
\Pi^\beta \; , \cr d\Pi^\alpha = 0 \; , \cr d\Pi^\prime_{\alpha} =
i \, {\Pi}_{\alpha\beta} \wedge \Pi^\beta \; ,
\end{cases}
\end{eqnarray}
where, for definiteness, we have set the inessential constant to
$\gamma_1= 1/64$.

We  can also write in this notation the MC equations of the
$\tilde {\mathfrak E}^{(528|32+32)}(s)$ superalgebra,
\begin{eqnarray}
\label{MC:s} && \tilde {\mathfrak E}^{(528|32+32)}(s) \quad :
\qquad
\nonumber \\
&& \qquad \begin{cases} d{\Pi}^{\alpha\beta} = -  i \Pi^\alpha
\wedge \Pi^\beta \; , \cr d\Pi^\alpha = 0 \; , \cr
d\Pi^\prime_{\alpha} = i \, ({\Pi}_{\alpha\beta} + s/32 \; \Pi^a
\Gamma_{a\alpha\beta} + s/32^{_.}6! \; \Pi^{a_1\ldots a_5}
\Gamma_{a_1\ldots a_5\alpha\beta})
 )\wedge
\Pi^\beta \; .
\end{cases}
\end{eqnarray}
For $s\not= 0$ the last equation in (\ref{MC:s}) involves
explicitly the $D$=11 gamma--matrices, so that $\tilde {\mathfrak
E}^{(528|32+32)}(s)$ possess only $SO(1,10)$ automorphisms.

Softening the $\tilde {\mathfrak E}^{(528|32+32)}(0)$
Maurer--Cartan equations by introducing  nonvanishing  curvatures
and a nontrivial spin connection, the Lie algebra is converted
into a gauge  FDA generated by $E^a$, $\psi^\alpha$,
$\omega^{ab}$, $B_1^{ab}$, $B_1^{a_1,\ldots,a_5}$ {\it and}
$\eta_{1 \alpha}$ and their curvatures (Eqs.~(\ref{CJS:Ta=}),
(\ref{CJS:Tf=}), (\ref{CJS:RL=}), (\ref{cBab}), (\ref{cB15}),
(\ref{cDeta})) which can be rewritten in terms of
\begin{eqnarray}
 \label{Epsi2}
& {\cal E}^{\alpha\beta}  \; , \qquad {\psi}^{\alpha} \; ,  \qquad
   \eta_{1 \alpha} \; ; \qquad \\ & \label{omaux} \omega^{ab} \; ,  \qquad
\end{eqnarray}
and their curvatures,
\begin{eqnarray}
\label{FD:TE} {\cal R}^{\alpha\beta} &=& D{\cal E}^{\alpha\beta}
+ i \psi^\alpha  \wedge \psi^\beta \; , \qquad \\
\label{FDpsi} \mathbf{R}^\alpha &=& D\psi^\alpha := d\psi^\alpha -
\psi^\beta \wedge \omega_\beta{}^\alpha \; ,
\\ \label{FDeta}
{\cal B}_{2\alpha } &=& D\eta_{1\alpha} - i \, {\cal
E}_{\alpha\beta} \wedge \psi^\beta \; ,
\end{eqnarray}
(setting again $\gamma_1=1/64$ in Eq. (\ref{FDeta})) plus
\begin{eqnarray}
 \label{FD:RL=}
\mathbf{R}^{ab} &:=& R^{ab} = d\omega^{ab} - \omega^{ac}\wedge
\omega_c{}^b \;  \quad
\end{eqnarray}
obeying the Bianchi identities (\ref{BI:Rff}), (\ref{BI:Rf}),
 \begin{eqnarray}\label{BI:Deta}
{\cal D} {\cal B}_{2\alpha } &:= & D{\cal B}_{2\alpha } + i \,
{\cal E}_{\alpha\beta} \wedge \mathbf{R}^\beta - i \, {\cal
R}_{\alpha\beta} \wedge \psi^\beta + {\cal R}_{\alpha}{}^{\beta}
\wedge \eta_{1 \beta} \equiv 0 \; ,
\end{eqnarray}
and (\ref{BI:RL}). The $\alpha\beta$ indices are rised and lowered
in (\ref{FDeta}) and (\ref{BI:Deta}) by the charge conjugation
matrix $C_{\alpha\beta}$ and its inverse $C^{\alpha\beta}$;
 notice that the gamma matrices now only appear in the spin
connection $\omega_{\alpha}{}^\beta = 1/4 \omega^{ab}\Gamma_{ab}
{}_{\alpha}{}^\beta$
 that enters in the covariant derivative $D$.
 Thus replacing formally the spin connection
$\omega_{\alpha}{}^\beta$ by a more general symplectic one
$\Omega_{\alpha}{}^\beta$ (restricted only by ${\cal
D}C_{\alpha\beta}=0$ which implies $\Omega_{\alpha\beta}:=
\Omega_{\alpha}{}^\gamma C_{\beta \gamma}= \Omega_{\beta\alpha}=
\Omega_{(\alpha\beta)}$) we arrive at a FDA possessing a local
$Sp(32,\mathbb{R})$ symmetry.

As discussed in \cite{Lett}, the superalgebra $\tilde {\mathfrak
E}(0)+\!\!\!\!\!\!\supset so(1,10)$, corresponding to the FDA
(\ref{Epsi2})--(\ref{FD:RL=}) without the replacement of
$\omega_{\alpha}{}^\beta$ by $\Omega_{\alpha}{}^\beta$, is an
expansion \cite{JdA02} of the supergroup $OSp(1|32)$, denoted
$OSp(1|32)(2,3,2)$, of dimension $647=583+64$; the superalgebra
$\tilde {\mathfrak E}(0)+\!\!\!\!\!\!\supset sp(32)$,
corresponding to the FDA (\ref{Epsi2})--(\ref{FD:RL=}) with
$\omega_{\alpha}{}^\beta$ replaced by the $sp(32)$ valued
$\Omega_{\alpha}{}^\beta$, is given by the expansion
$OSp(1|32)(3,2)$ of dimension $1120=1056+64$. We refer to
\cite{JdA02,Lett} for details.

\subsection{Composite nature of the $A_3$
three-form gauge field}

 The problem \cite{D'A+F} now is to express the form $A_3$ defined
by the CJS FDA relations (\ref{CJS:R4=}), (\ref{DF4}) [with
(\ref{CJS:Ta=})--(\ref{CJS:RL=}) (\ref{BITa})--(\ref{DR})] in
terms of the one--forms $B_1^{ab}$, $B_1^{abcde}$, $\eta_{1
\alpha}$ plus the original  graviton and gravitino one--forms,
$E^a$ and $\psi^\alpha$. For it, we write the most general
expression for a three-form $A_3$ in terms of the above one-forms,
\begin{eqnarray}
\label{A3=Ans} A_3 &=& {\lambda\over 4}  B_1^{ab} \wedge E_a
\wedge E_b \; - { \alpha_1\over 4} B_{1\, ab} \wedge B_1^b{}_c
\wedge B_1^{ca} - {\alpha_2\over 4} B_{1\, b_1a_1\ldots a_4}
\wedge B_1^{b_1}{}_{b_2} \wedge B_1^{b_2a_1\ldots a_4} -
\nonumber \\
&-& {\alpha_3\over 4}  \epsilon_{a_1\ldots a_5b_1\ldots b_5c}
B_1^{a_1\ldots a_5} \wedge B_1^{b_1\ldots b_5} \wedge E^c -
\nonumber \\
&-& {\alpha_4\over 4}  \epsilon_{a_1\ldots a_6b_1\ldots b_5}
B_1^{a_1a_2 a_3}{}_{c_1c_2} \wedge B_1^{a_4a_5a_6c_1c_2} \wedge
B_1^{b_1\ldots b_5} -
\nonumber \\
&-& {i\over 2} \psi^\beta \wedge \eta_1^\alpha \wedge \left(
\beta_1 \, E^a \Gamma_{a\alpha\beta} -i  \beta_2 \, B_1^{ab}
\Gamma_{ab\; \alpha\beta} + \beta_3 \, B_1^{abcde} \Gamma_{abcde\;
\alpha\beta} \right) \; ,
\end{eqnarray}
 and look for the values of the constants $\alpha_1, \ldots,
\alpha_4$, $\beta_1, \ldots, \beta_3$ {\it and} $\lambda$ such
that $A_3$ of Eq. (\ref{A3=Ans}) obeys (\ref{CJS:R4=}) {\it for
arbitrary curvatures} of the one--form fields. The numerical
factors  in the right hand side of (\ref{A3=Ans}) are introduced
to make the definition of the coefficients coincide with that in
\cite{D'A+F} while keeping our notation for the FDA and
supergravity constraints. The only essential difference with
\cite{D'A+F} is the inclusion of the arbitrary coefficient
$\lambda$ in the first term; as we  show below this leads to a
one-parametric family of solutions that includes  the two
D'Auria--Fr\'e ones.

Factoring out the coefficients for the various independent forms
one finds a system of equations given by \footnote{The factor $5!$
in (\ref{Eq5}) is missing in footnote 6 in \cite{Lett}.}
\begin{eqnarray}\label{Eq0}
\beta_1 + 10\beta_2 - 6! \beta_3=0\; ,
 \end{eqnarray}
and
\begin{eqnarray}
\label{Eq1} \lambda - 2\delta\beta_1=1 \; ,
\\
\label{Eq2} \lambda- 2\gamma_1 \beta_1 - 2\delta\beta_2 =0 \; ,
\\
\label{Eq3} 3\alpha_1 + 8 \gamma_1\beta_2 =0 \; ,
\\
\label{Eq4} \alpha_2 - 10\gamma_1 \beta_3 - 10\gamma_2 \beta_2 =0
\; , \\ \label{Eq5} 5! \alpha_3 - \delta \beta_3 - \gamma_2
\beta_1
=0 \; , \\
\label{Eq6} \alpha_2 - 5!\,  10\gamma_2 \beta_3
=0 \; , \\
\label{Eq7} \alpha_3 - 2\gamma_2 \beta_3
=0 \; , \\
\label{Eq8} 3 \alpha_4 + 10\gamma_2 \beta_3 =0 \; .
\end{eqnarray}
Eq. (\ref{Eq0}) comes from the cancellation of terms proportional
to the $\psi\wedge \psi \wedge \psi \wedge \eta_1$ form in Eq.
(\ref{CJS:R4=}) with (\ref{A3=Ans}); Eqs. (\ref{Eq1})--(\ref{Eq8})
from the cancellation of terms proportional to $\psi\wedge \psi$
times different products of bosonic forms $E^a$, $B_1^{ab}$,
$B_1^{abcde}$ [namely, $\psi^\alpha \wedge \psi^\beta \wedge E^a
\wedge E^b \Gamma_{ab\alpha\beta}$ for (\ref{Eq1}); $\psi^\alpha
\wedge \psi^\beta \wedge B_1^{ab} \wedge E_b
\Gamma_{a\alpha\beta}$ for (\ref{Eq2}); {\it etc.}].

For the nonvanishing curvatures of the one--form fields, Eq.
(\ref{CJS:R4=})  with (\ref{A3=Ans}) gives also the expression for
the four--form curvature $\mathbf{R}_4$ in terms of the curvatures
of the one--forms; setting ${\mathbf{R}}^a=0$ (which is proper in
the description of supergravity constraints as well as for  the
component formulation of supergravity with ``supersymmetric spin
connections'') one gets
 \begin{eqnarray}
\label{R4=Ans} \mathbf{R}_4 &=& {\lambda\over 4} {\cal B}_2^{ab}
\wedge E_a \wedge E_b \; - {3 \alpha_1\over 4} {\cal B}_{2 ab}
\wedge B_1^b{}_c \wedge B_1^{ca} - {\alpha_2\over 2} {\cal B}_{2\,
a_1\ldots a_5} \wedge B_1^{a_1}{}_{b} \wedge B_1^{ba_2\ldots a_5}
+ \nonumber \\
&+& { \alpha_2\over 4} B_{1\, a_1\ldots a_5} \wedge {\cal
B}_2^{a_1}{}_{b} \wedge B_1^{ba_2\ldots a_5} -
 {\alpha_3\over 2}  \epsilon_{a_1\ldots a_5b_1\ldots b_5c}
E^c \wedge B_1^{a_1\ldots a_5} \wedge  {\cal B}_2^{b_1\ldots b_5}
- \nonumber \\
&-&  {\alpha_4 \over 4} \epsilon_{a_1\ldots a_6b_1\ldots b_5}
B_1^{a_1a_2 a_3}{}_{c_1c_2} \wedge B_1^{a_4a_5a_6c_1c_2} \wedge
{\cal B}_2^{b_1\ldots b_5} - \nonumber \\  &-& {\alpha_4\over 2}
\epsilon_{a_1\ldots a_6b_1\ldots b_5} B_1^{a_4a_5a_6c_1c_2} \wedge
B_1^{b_1\ldots b_5} \wedge {\cal B}_2^{a_1a_2 a_3}{}_{c_1c_2} -
\nonumber \\
&-& {i\over 2}  \psi^\beta \wedge \eta_1^\alpha \wedge \left(-i
\beta_2 \, {\cal B}_2^{ab} \Gamma_{ab\; \alpha\beta} + \beta_3 \,
{\cal
B}_2^{abcde} \Gamma_{abcde\; \alpha\beta} \right) +\nonumber \\
&+& {i\over 2} \psi^\beta \wedge \left( \beta_1 \, E^a
\Gamma_{a\alpha\beta} -i  \beta_2 \, B_1^{ab} \Gamma_{ab\;
\alpha\beta} + \beta_3 \, B_1^{abcde} \Gamma_{abcde\; \alpha\beta}
\right) \wedge {\cal B}_2^\alpha + \nonumber \\  &+& {i\over 2}
\eta^\alpha \wedge \left( \beta_1 \, E^a \Gamma_{a\alpha\beta} -i
\beta_2 \, B_1^{ab} \Gamma_{ab\; \alpha\beta} + \beta_3 \,
B_1^{abcde} \Gamma_{abcde\; \alpha\beta} \right) \wedge
\mathbf{R}^\beta \; .
\end{eqnarray}
Eq.~(\ref{R4=Ans}) assumes that the relations
(\ref{Eq0})--(\ref{Eq8}) among the coefficients $\lambda,
\alpha_1, \ldots$ are satisfied. These equations, actually
necessary conditions for a composite structure of the three-form
$A_3$, also solve the problem of trivializing the cocycle
(\ref{R4=0})  on a suitable flat (or rigid) enlarged superspace to
which we now turn.

\subsection{Trivializing the four-cocycle $dA_3$,
enlarged superspaces,  and  the \\ fields/extended superspace
variables correspondence }
  \label{adboscoor}

The trivialization of the $w_4\; $ ($dA_3$) standard supersymmetry
algebra four-cocycle on a larger Lie superalgebra implies
expressing the form $A_3$ obeying (\ref{R4=0}) in terms of  MC
one--forms on a larger supergroup manifold,  {\it i.e.} on a {\it
generalized superspace} (see \cite{JdA00}) with additional
coordinates. The above described calculations for the case of
vanishing curvatures leading to Eqs. (\ref{Eq0})--(\ref{Eq8})
considered $E^a$, $\psi^{\alpha}$, $B_1^{ab}$, $B_1^{abcde}$,
$\eta_{1\,\alpha}$ as independent one-forms.  This is tantamount
to looking for a trivialization of the four-cocycle $w_4$ on an
extended superalgebra $\tilde{\mathfrak{E}}^{(528|32+32)}$
associated to the {\it rigid} superspace (group manifold)
$\tilde{\Sigma}^{(528|32+32)}$
 with
$517=55+462$ bosonic and $32$ fermionic {\it additional}
coordinates.

The original $32$ fermionic coordinates $\theta^\alpha$ are
associated with the fermionic one--forms $\psi^\alpha$, which by
Eq.~(\ref{CJS:Tf=}) become closed, $d\psi^\alpha=0$,  for vanishing
curvature $\mathbf{R}^\alpha=0$ and trivial spin connection
$\omega^{ab}=0$;  as a closed form, $\psi^\alpha = d\theta^\alpha
\equiv \Pi^\alpha$. Similarly, $\mathbf{R}^a=0$, now reads $dE^a +
id\theta^{\alpha }\wedge d\theta^{\beta }\Gamma^a_{\alpha\beta} =0$
and has the invariant solution $E^a=\Pi^a = dx^a -id\theta^{\alpha }
\Gamma^a_{\alpha\beta} \theta^{\beta }$ on the standard superspace
$\Sigma^{(11|32)}$ of coordinates $Z=(x^\mu , \theta^\alpha)$. On
$\Sigma^{(11|32)}$ all other differential forms, {\it e.g.}
$B^{ab}_1$ and $B^{a_1\ldots a_5}_1$, can be expressed (e.g.,
$B^{ab}_1=E^d B_d^{ab} (Z) + E^\gamma B_\gamma^{ab} (Z)$) in the
basis provided by $E^a= \Pi^a$ and $\psi^\alpha= \Pi^\alpha$. This
is true also for the curved standard superspace,  but in this case
$E^a$ and $\psi^\alpha= E^\alpha$ are `soft' one-forms and not the
invariant MC forms $\Pi^a$ and $\Pi^\alpha$.

If one considers the bosonic differential one-forms $B^{ab}_1$ and
$B^{a_1\ldots a_5}_ 1$ as independent, one implicitly assumes that
$dB^{ab}_1= - d\theta^\alpha \wedge d\theta^\beta
\Gamma^{ab}_{\alpha\beta}$ and $dB^{a_1\ldots a_5}_1= - i
d\theta^\alpha \wedge d\theta^\beta \Gamma^{a_1\ldots
a_5}_{\alpha\beta}$ (see Eqs. (\ref{dB2=}), (\ref{dB5=}))  may be
solved in terms of invariant one-forms
\begin{eqnarray}\label{B2=Pi2}
B^{ab}_1= \Pi^{ab}: = dy^{ab}- d\theta^\alpha
\Gamma^{ab}_{\alpha\beta}\theta^\beta \; , \quad  \nonumber \\
B^{a_1\ldots a_5}_1= \Pi^{a_1\ldots a_5} := dy^{a_1\ldots a_5} - i
d\theta^\alpha \Gamma^{a_1\ldots a_5}_{\alpha\beta}\theta^\beta \;.
\end{eqnarray}
The new parameters $y^{ab}$ and $y^{a_1\ldots a_5}$ entering in
$B^{a_1 a_ 2}_1$ and $B^{a_1\ldots a_5}_1$ constitute the
additional $(55+462=517)$ bosonic variables of an extended
superspace. This is the extended $\Sigma^{(528|32)}$ rigid
superspace, which may be considered as a supergroup for which
$\Pi^a$, $\Pi^{a_1 a_2}$, $\Pi^{a_1,\ldots,a_5}$ and $\Pi^\alpha$
are all invariant MC forms\footnote{The $\Sigma^{(528|32)}$
extended superspace group may be found in our spirit by searching
for a trivialization of the $\mathbb{R}^{528}$--valued two-cocycle
$d{\cal E}^{\alpha\beta}=-id\theta^{\alpha }\wedge
d\theta^{\beta}$, which leads to the one-form
 ${\cal E}^{\alpha\beta}=dX^{\alpha\beta} -id\theta^{(\alpha}\theta^{\beta)}$.
 This introduces in a natural way the 528 bosonic coordinates
$X^{\alpha \beta}$ and the transformation law $\delta_\epsilon
X^{\alpha \beta} = i \theta^{(\alpha} \epsilon^{\beta)}$ that
makes ${\cal E}^{\alpha \beta}$ invariant, and hence leads to a
central extension structure for the extended superspace group
$\Sigma^{(528|32)}$, which is parametrized by $(X^{\alpha\beta},
\theta^{\alpha})$. The 528 bosonic coordinates include, besides
the
 standard spacetime $x^\mu=1/32\, X^{\alpha\beta}\Gamma^\mu_{\alpha\beta}$
 ones, $517=55+462$ tensorial additional coordinates,
 $y^{\mu\nu}=1/64!X^{\alpha\beta}\Gamma^{\mu\nu}_{\alpha\beta}$
 plus $y^{\mu_1\ldots \mu_5}= 1/64!X^{\alpha\beta}\Gamma^{\mu_1\ldots
\mu_5} _{\alpha\beta}$. The (maximally extended in the bosonic
sector) superspace $\Sigma^{(528|32)}$ transformations make of
${\cal E}^{\alpha \beta}$ a MC form that trivializes, on the
extended superalgebra ${\mathfrak{E}}^{(528|32+32)}$, the
non-trivial CE two-cocycle on the original odd abelian algebra
$\Sigma^{(0|32)}$.}. When the curvatures are not zero, and in
particular ${\cal B}_2^{ab} \neq 0$, ${\cal B}_2^{a_1\ldots a_5}
\neq 0$ {\it i.e.}, the invariant one-forms $B^{a_1 a_ 2}_1$,
$B^{a_1\ldots a_5}_1$ become `soft',  $\Sigma^{(528|32)}$  is non flat
and no longer a group manifold.

The additional fermionic form $\eta_{1 \alpha}$ for the case of
vanishing curvatures obeys (see Eqs. (\ref{deta=}), (\ref{MC:s}))
\begin{eqnarray}
 \label{cDeta=0}
d\eta_{1\alpha} &=& i \, \left(\delta \, E^a\Gamma_a - i \gamma_1
B_1^{ab} \Gamma_{ab} +  \gamma_2  B_1^{a_1\ldots a_5}
\Gamma_{a_1\ldots a_5} \right) _{\alpha\beta} \wedge \psi^\beta \;
.
\end{eqnarray}
The two-form on the {\it r.h.s.}~of this equation is a nontrivial
two-cocycle on ${\mathfrak{E}}^{(528|32)}$. It may be trivialized
by adding $32$ new fermionic coordinates $\theta^{\prime}_\alpha$
that are used to solve
  (\ref{cDeta=0}) by
\begin{eqnarray}\label{eta=dt}
\eta_{1\alpha} &=& \Pi^\prime_{\alpha} := d\theta^{\prime}_\alpha
+ i \left(\delta \, E^a\Gamma_a - i \gamma_1  B_1^{ab} \Gamma_{ab}
  + \gamma_2  B_1^{a_1\ldots a_5} \Gamma_{a_1\ldots a_5} \right)_{\alpha\beta}
\theta^\beta    \qquad \nonumber \\
&-& {2\over 3}  \delta \, d\theta\Gamma^a\theta \,
(\Gamma_a\theta)_\alpha + {2\over 3} \gamma_1
d\theta\Gamma^{ab}\theta \, (\Gamma_{ab}\theta)_\alpha - {2\over
3} \gamma_2 d\theta\Gamma^{a_1\ldots a_5}\theta \,
(\Gamma_{a_1\ldots a_5}\theta)_\alpha \; .
\end{eqnarray}
In the next section  we will show  that  the trivialization of the
CE cocycles encoded in the FDA of Eqs. (\ref{CJS:Ta=}),
(\ref{CJS:Tf=}), (\ref{cBab})-(\ref{cDeta}) is possible on all the
extended superalgebras ${\mathfrak{E}}^{(528|32+32)}(s\not=0)$
associated with the superspace groups $\Sigma^{(528|32+32)}(s)$
parametrized by
\begin{eqnarray}\label{32+32}
 \Sigma^{(528|32+32)}(s) \; : \;
 \left( x^\mu \; , \; y^{\mu\nu} \; , \; y^{\mu_1\ldots \mu_5}\; ;
\; \theta^\alpha \; , \; \theta^\prime_\alpha \right)  \; ,
\end{eqnarray}
where the $\theta^\prime_\alpha$ coordinate (the `second' 32,
corresponding to  the fermionic central charge) is associated with
the one-form $\eta_{1 \alpha}$ through (\ref{eta=dt}).

The softening of the  ${\mathfrak{E}}^{(528|32+32)}$ MC equations
leads to the associated  gauge FDA, with as many one--form gauge
fields as group parameters. This is one more example of the
fields/extended superspace coordinates correspondence already
mentioned (See also Sec.~6).

\subsection{Underlying gauge superalgebras for $D=11$ supergravity and
their associated $A_3$ composite structures}

As it was discussed above, the constants $\delta$, $\gamma_1$,
$\gamma_2$ restricted by Eq.~(\ref{id1}) or, equivalently,
expressed through the parameter $s$ by (\ref{s-def}), determine
the superalgebras
$\tilde{\mathfrak{E}}(s)=\tilde{\mathfrak{E}}^{(528|32+32)}(s)$
that are not isomorphic.

On the other hand, these constants appear in the system of equations
(\ref{Eq0})--(\ref{Eq8}) as parameters. Clearly, as we have found
nine equations for eight constants $\alpha_1, \ldots ,\alpha_4$,
$\beta_1, \beta_2, \beta_3$, {\it and} $\lambda$, the existence of
solutions for the system of equations (\ref{Eq0})--(\ref{Eq8}) is
not guaranteed for arbitrary values of $\delta$, $\gamma_1$,
$\gamma_2$ obeying (\ref{id1}), {\it i.e.} for
$\tilde{\mathfrak{E}}(s)$ with an arbitrary $s$. One might expect to
have one more condition on these constants that would fix them
completely up to a rescaling of the `new' fermionic form $\eta_{1
\alpha}$ {\it i.e.}, that would fix completely the parameter $s$ in
(\ref{s-def}) and, hence, would select only {\it one} representative
of the $\tilde{\mathfrak{E}}^{(528|32+32)}(s)$ family of
superalgebras. However, already the existence of {\it two} solutions
\cite{D'A+F} of the more restricted system (\ref{Eq0})--(\ref{Eq8})
for $\lambda=1$, indicates that this is not the case. Indeed, the
system (\ref{Eq0})--(\ref{Eq8}) contains dependent equations. These
may be identified as the equations for the constants $\beta_2$ and
$\beta_3$ that come from the two equations for $\alpha_2$, Eqs.
(\ref{Eq4}) and (\ref{Eq6}),
\begin{eqnarray}\label{b2b3=1}
\gamma_2 \beta_2 + (\gamma_1 -  5!\, \gamma_2) \beta_3 =0 \; ,
\end{eqnarray}
 and the equation
\begin{eqnarray}\label{b2b3=2}
10 \gamma_2 \beta_2 - (\delta +  4\cdot5!\, \gamma_2) \beta_3 =0
\; ,
\end{eqnarray}
obtained from the two expressions with $\alpha_3$, Eqs.
(\ref{Eq5}) and (\ref{Eq7}), after $\beta_1$ is removed by means
of  (\ref{Eq0}). One can easily check that Eq. (\ref{b2b3=2})
coincides with (\ref{b2b3=1}) as far as $\delta$, $\gamma_1$ and
$\gamma_2$ obey (\ref{id1}).

Thus the general solution of the system of Eqs.
(\ref{Eq0})--(\ref{Eq8}) plus (\ref{id1}) is effectively
one-parametric. It may be given in terms of two parameters
$\delta$ and $\gamma_1$,
\begin{eqnarray}\label{sEq=g}
\beta_1 & = & {2\over 5}\; {5 \gamma_1-  \delta \; \over (2
\gamma_1-  \delta)^2} \; , \quad \nonumber \\  \quad \beta_2 & = &
{1\over 10}\; {4\gamma_1 +\delta \; \over (2 \gamma_1-  \delta)^2}
\; , \quad
\nonumber \\
 \beta_3  & = & {1\over 10^. 5!} \; {{10\gamma_1 +\delta}  \over   (2 \gamma_1-  \delta)^2}
\; , \quad \nonumber \\
\alpha_1 & = &  -
{8\over 3} \gamma_1\beta_2 = - {4\over 15}\; {\gamma_1
(4\gamma_1+\delta) \; \over (2 \gamma_1- \delta)^2}
\; ,  \nonumber \\
\alpha_2 & = & 10\; 5! \; \gamma_2\beta_3 = {1\over 6!} \;
{(10\gamma_1+\delta)^2 \over (2 \gamma_1-  \delta)^2}
\; , \quad   \nonumber \\
\alpha_3  & = & 2 \gamma_2\beta_3 = {1\over 5^. 6!\, 5!} \; {
{(10\gamma_1 +\delta)^2} \over   (2 \gamma_1- \delta)^2} \; ,
\quad
\nonumber \\
 \alpha_4 &=& - {10\over 9} \gamma_2\beta_3 = - {1\over 9^. 6!\, 5!} \;
 { {(10\gamma_1 +\delta)^2}  \over   (2 \gamma_1-  \delta)^2}
\; , \quad
\nonumber \\
\lambda &=& 1 + 2\delta \beta_1= {1 \over 5} \; { {20\gamma^2_1
+\delta^2} \over
 (2 \gamma_1-  \delta)^2} \; .
\end{eqnarray}
However, as the value of one parameter ($\delta$ if nonvanishing,
$\gamma_1$ otherwise) can be used to rescale the new fermionic
form $\eta_{1 \alpha}$ we see that, effectively, there is a {\it
one-parameter family of solutions}. This indicates that the
trivialization of the four--cocycle is possible on (almost all)
the enlarged supergroup manifolds
$\tilde{\Sigma}^{(528|32+32)}(s)$ associated with the
superalgebras $\tilde{\mathfrak{E}}^{(528|32+32)}(s)$.

To find the singular points, let us write the general solution
(\ref{sEq=g}) in terms of the parameter $s$ as defined in Eq.
(\ref{s-def}),
\begin{eqnarray}\label{sEq=g(s)}
 \tilde{\mathfrak{E}}^{(528|32+32)}(s)\; : \hspace{1cm} && \; \nonumber \\  &&
 \delta=2\gamma_1 (s+1) \; , \qquad \gamma_2 =2\gamma_1({s \over 6!} + {1 \over 5!}) \; ;  \nonumber \\
&& \lambda = {1 \over 5} \; \frac{s^2+2s+6}{s^2} \; , \qquad
\nonumber \\
\beta_1 &=&-\frac{1}{10\gamma_1} \; \frac{2s-3}{s^2} \; , \qquad
\beta_2=\frac{1}{20\gamma_1} \; \frac{s+3}{s^2} \; , \qquad
\nonumber \\  \beta_3&=&\frac{3}{10 \cdot 6! \gamma_1} \;
\frac{s+6}{s^2}  \; ,
\nonumber \\
 \alpha_1 &=&-\frac{1}{15} \; \frac{2s+6}{s^2} \; ,  \qquad
\alpha_2=\frac{1}{6!} \; \frac{(s+6)^2}{s^2} \; , \qquad \nonumber
\\  \alpha_3&=&\frac{1}{5 \cdot 6! 5!} \; \frac{(s+6)^2}{s^2} \; ,
\quad \alpha_4=- \frac{1}{9 \cdot 6! 5!} \; \frac{(s+6)^2}{s^2} \;
. \quad
\end{eqnarray}
We see that the only forbidden value  is $s=0$. Thus Eqs.
(\ref{sEq=g(s)}) or, equivalently, (\ref{sEq=g}), trivialize the
$\tilde{\mathfrak{E}}^{(11|32)}$  CE four--cocycle
 on the extended $\tilde{\mathfrak{E}}^{(528|32+32)}(s\not=0)$
 superalgebra,  with associated supergroup manifold
$\tilde{\Sigma}^{(528|32+32)}(s\not=0)$. The same Eqs.
(\ref{sEq=g(s)}) (or (\ref{sEq=g})) determine the  $A_3$
three--form gauge field by (\ref{A3=Ans})  in terms of the
one--form gauge fields of the
$\tilde{\mathfrak{E}}^{(528|32+32)}(s\not=0)$ superalgebra; these
make $A_3$ a composite, rather than a fundamental field. We stress
once more that the values of $\delta$, $\gamma_1$ and $\gamma_2$
determine the Lie superalgebras
$\tilde{\mathfrak{E}}^{(528|32+32)}(s)$ associated with
$\Sigma^{(528|32+32)}(s)$, while those of
$\alpha_1,\ldots,\alpha_4$ and $\beta_1,\ldots,\beta_3$ determine
the expression of $A_3$ in Eq. (\ref{A3=Ans}) (the trivialization
of the cocycle).

Setting $\lambda=1$, as in \cite{D'A+F}, we find only {\it two}
solutions, one corresponding to $s=3/2$
\begin{eqnarray}\label{D'A+F1}
\tilde{\mathfrak{E}}^{(528|32+32)}(3/2)\;  : & \nonumber \\
 & {} \;  \delta = 5\gamma_1 \not=0 \; ,
 \quad \gamma_2=  {\gamma_1 \over 2\, 4!} \; ;
\nonumber \\
& \lambda=1 \; , \qquad  {}  \nonumber
\\ & \beta_1=0\; , \qquad \beta_2 = {1\over 10\gamma_1 }
\; ,  \qquad
\beta_3 = {1\over 6! \, \gamma_1} \; , \nonumber \\
& \alpha_1= - {4\over 15}\; , \quad  \alpha_2= {25\over 6!}  \; ,
\quad  \alpha_3= {1\over 6!\, 4! } \; , \quad
\alpha_4= - {1\over 54\, (4!)^2}  \; ,  \nonumber \\
\end{eqnarray}
 and a second corresponding to $s=-1$,
\begin{eqnarray}\label{D'A+F2}
\tilde{\mathfrak{E}}^{(528|32+32)}(-1)\;  : \qquad {}\quad & \nonumber \\
  &  \delta  =0 \; ,  \quad
 \gamma_2=  {1\over 3^. 4!} \; ,\quad \gamma_1 \not= 0 \; ;
\nonumber \\ & \lambda=1 \; , \qquad  {}  \nonumber
\\ & \beta_1={1\over 2 \gamma_1 } , \qquad \beta_2 =
{1\over 10\, \gamma_1 } ,
\quad  \beta_3 = {1\over 4^. 5! \gamma_1 }, \nonumber \\
& \alpha_1=  - {4\over 15}\; , \quad  \nonumber
\\ & \alpha_2=  {25 \over 6!} \;
, \quad \alpha_3=   {1\over 6! \, 4!}\; , \quad \alpha_4=  -
{1\over 54\, (4!)^2}\; . \quad
\end{eqnarray}
Notice that the values of $\alpha_{1,2,3,4}$ are the same for
both of them. The two D'Auria and Fr\'e solutions correspond to
$\delta=1$ in Eq. (\ref{D'A+F1}) and $\gamma_1=-1/2$ in Eq.
(\ref{D'A+F2}).

Due to the presence of $\lambda$, we have more possibilities. A
particularly interesting solution of our system is found by
setting $s=-6$ in (\ref{sEq=g(s)}) or $\gamma_2=
\frac{1}{6!}(10\gamma_1 + \delta)=0$ in (\ref{sEq=g}):
\begin{eqnarray}\label{minimal}
\tilde{\mathfrak{E}}^{(528|32+32)}(-6) \; : \qquad & \nonumber \\
 & \delta
=-10\gamma_1 \; ,  \quad
 \gamma_2=  0 \; ; \nonumber \\ & \lambda={1 \over 6} \; , \qquad
\nonumber \\
& \beta_1={1\over 4! \gamma_1 } , \qquad \beta_2 = - {1\over 2\,
5!\, \gamma_1 } ,
\quad  \beta_3 = 0 , \nonumber \\
& \alpha_1=   {1\over 90}\; , \quad  \alpha_2= 0  \; , \quad
\alpha_3=  0 \; , \quad \alpha_4=  0 \; . \quad
\end{eqnarray}
This corresponds to an especially simple expression for the
composite $A_3$ form in terms of the gauge fields of this
$\tilde{\mathfrak{E}}^{(528|32+32)}(-6)$ superalgebra,
\begin{eqnarray}
\label{A3=min} A_3 &=&  {1\over 4!} B_1^{ab} \wedge E_a \wedge E_b
\; - {1\over 3^.5!}
 B_{1\, ab} \wedge B_1^a{}_c \wedge
B_1^{cb} - {i \over
5! \, 4 \gamma_1}  \psi^\beta \wedge \eta^\alpha \wedge \left( 10
\, E^a \Gamma_{a\alpha\beta} + i \, B_1^{ab}
\Gamma_{ab\;\alpha\beta}
\right) \;  \nonumber \\
\end{eqnarray}
and to a shorter version of the additional  fermionic curvature
(\ref{cDeta})
\begin{eqnarray}
\label{Deta=min} {\cal B}_{2\alpha } &=& D\eta_{1\alpha} + i \,
\gamma_1 ( 10 E^a \Gamma_{a} + i B_1^{ab}
\Gamma_{ab})_{\alpha\beta} \wedge \psi^\beta \; . \qquad \quad
\end{eqnarray}
The $\gamma_2=0$ choice of Eq.~(\ref{minimal}) implies that the
underlying Lie superalgebra
$\tilde{\mathfrak{E}}^{(528|32+32)}(-6)$  includes $Z_{a_1\ldots
a_5}$ also as central generator ({\it cf.} last line in
Eq.~(\ref{PQ})),
\begin{eqnarray}\label{S(-6)}
\tilde{\mathfrak{E}}^{(528|32+32)}(-6) \; : \hspace{2cm} && \nonumber \\
\{ Q_\alpha , Q_\beta \} & = & P_a \Gamma^a_{\alpha\beta} + Z_{ab}
i\Gamma^{ab}_{\alpha\beta} + Z_{a_1\ldots a_5} \Gamma^{a_1\ldots
a_5 }_{\alpha\beta} \; , \nonumber \\ {} [ P_a , Q_\alpha ] &=& -
10\gamma_1 \; \Gamma_{a \; \alpha\beta} Q^{\prime\beta} \; , \quad
\nonumber
\\ {} [ Z_{a_1a_2} , Q_\alpha ] &=& i\gamma_1
\Gamma_{a_1a_2 \; \alpha\beta} Q^{\prime\beta} \; , \nonumber \\
&& {}  [ Z_{a_1 \ldots a_5} , \, all\;  ]= 0 \; , \nonumber \\
&& {}  [ Q^{\prime\alpha}  ,  \, all\;  \} = 0 \; .
\end{eqnarray}
Thus one can consistently truncate
$\tilde{\mathfrak{E}}^{(528|32+32)}(-6)$  by setting the central
generator   $Z_{a_1\ldots a_5}$ equal to zero. In such a way one
arrives at the $\tilde {\mathfrak E}_{min}=\tilde {\mathfrak
E}^{(66|32+32)}$ superalgebra,  whose extension by $Z_{a_1\ldots
a_5}$ gives $\tilde {\mathfrak E}^{(528|32+32)}(-6)$ in
Eq.~(\ref{S(-6)}). Explicitly, $\tilde {\mathfrak E}_{min}$ is the
$(66+64)$-dimensional superalgebra
\begin{eqnarray}\label{susymin}
\tilde {\mathfrak
E}_{min}=\tilde{\mathfrak{E}}^{(66|32+32)}\; : \hspace{2cm} & & \nonumber \\
\{Q_\alpha,Q_\beta\} &=& \Gamma^a_{\alpha\beta} P_a +
i\Gamma^{a_1a_2}_{\alpha\beta} Z_{a_1a_2} \; , \nonumber \\ {} [
P_a , Q_\alpha ] &=&  -10 \gamma_1 \; \Gamma_{a \; \alpha\beta}
Q^{\prime\beta} \; , \qquad \nonumber \\ {}[ Z_{a_1a_2} ,
Q_\alpha]
&=& i \gamma_1 \Gamma_{a_1a_2 \; \alpha\beta} Q^{\prime\beta}\; , \nonumber \\
 && {}  [ Q^{\prime\alpha}  ,  \, all\;  \} = 0 \; .
\end{eqnarray}
that corresponds to the most economic enlargement of the standard
supersymmetry algebra $\tilde{\mathfrak{E}}^{(11|32)}$ for which the
$w_4$ four--cocycle (corresponding to $dA_3$, Eq. (\ref{R4=0}))
becomes trivial. Obviously,  the gauge FDA associated to
$\tilde{\mathfrak{E}}_{min}=\tilde{\mathfrak{E}}^{(66|32+32)}$ does
not involve $B_1^{a_1\ldots a_5}$ (see next Section).

The superalgebras $\tilde{\mathfrak{E}}^{(528|32+32)}(s\not=0)$
account for an {\it underlying gauge symmetry of the $D=11$
supergravity} in the sense that such a symmetry is hidden in the
original CJS formulation and only becomes explicit when  $A_3$ is
written in terms of the one--form gauge fields of
$\tilde{\mathfrak{E}}^{(528|32+32)}(s)$, $s\not=0$. These
superalgebras  may be considered themselves as nontrivial
deformations of $\tilde{\mathfrak{E}}^{(528|32+32)}(0)$.

The $s=0$ value (Eq. (\ref{sol1})) corresponds to the Lie
superalgebra $\tilde{\mathfrak{E}}^{(528|32+32)}(0)$ associated
with the superspace group $\tilde{\Sigma}^{(528|32+32)}(0)$. This
possesses $Sp(32)$ as its automorphism group, which is  not
allowed when  $s\not=0$, Eqs. (\ref{sEq=g(s)}) or (\ref{sEq=g}).
The full $\tilde{\Sigma}^{(528|32+32)}(0)\times\!\!\!\!\!\!\supset
Sp(32)$ group is isomorphic \cite{Lett} to the expansion
\cite{JdA02} $OSp(1|32)(2,3)$ of the $OSp(1|32)$ supergroup. If
the Lorentz connection is taken into account, the complete
symmetry group reduces to
$\tilde{\Sigma}^{(528|32+32)}(0)\times\!\!\!\!\!\!\supset
SO(1,10)$ which is isomorphic to the expansion $OSp(1|32)(2,3,2)$.
However, $\tilde{\Sigma}^{(528|32+32)}(0)$ does not allow for a
trivialization of the $w_4$ cocycle.  Equivalently, the problem of
the composite structure of $A_3$ form does not have a solution in
terms of $\tilde{\mathfrak{E}}^{(528|32+32)}(0)$ gauge fields.
This implies that the four-cocycle $w_4$ in Eq.~(\ref{R4=0}) (with
Lorentz rather than $Sp(32)$ invariance) may be trivialized only
on the superalgebras $\mathfrak{E}^{(528|32+32)}(s\not=0)$ with
$SO(1,10)$ automorphism group; the superspace
$\Sigma^{(528|32+32)}(0)$, with $Sp(32)$ automorphisms, on which
the new fermionic Cartan form could be given by
 $\eta_{1\alpha} = d\theta^{\prime}_\alpha + i
{\Pi}_{\alpha\beta}\,\theta^\beta - {2\over 3}
d\theta_{(\alpha}\theta_{\beta )}\, \theta^\beta$ ({\it cf.}
(\ref{eta=dt}) for (\ref{sol1}), see (\ref{cEff=def})),  does not
allow for such a trivialization.  This result is in a way natural:
if there should be a singularity  in the solution of the cocycle
trivialization conditions, it should be associated with an algebra
having particular properties. This is the case for that determined
by Eq. (\ref{sol1}) since only for these values of the constants
the rigid superspace group $\Sigma^{(528|32+32)}(0)$
 automorphism symmetry is enhanced from $SO(1,10)$
to $Sp(32)$.

\subsection{An economic underlying gauge group
structure for $D=11$ supergravity and the generalized superspace
$\Sigma^{(66|64)}$}

Our analysis has shown that the minimal FDA allowing for a
composite structure of the CJS 3--form $A_3$ can be defined,
fixing $\gamma_1=-1$,  by
\begin{eqnarray}\label{min:Ta=}
\mathbf{R}^a &=& DE^a + i \psi^\alpha \wedge \psi^\beta
\Gamma^a_{\alpha\beta} \; ,
\\ \label{min:Tf=}
\mathbf{R}^\alpha &:=& D\psi^\alpha :=  d\psi^\alpha - \psi^\beta
\wedge \omega_\beta{}^\alpha \; .
\\ \label{min:RL=}
\mathbf{R}^{ab} &:=& R^{ab}=d\omega^{ab} - \omega^{ac}\wedge
\omega_c{}^b \; , \\
\label{minBab} {\cal B}_2^{ab} &=& DB_1^{ab} + \psi^\alpha  \wedge
\psi^\beta \, \Gamma^{ab}_{\alpha\beta} \; , \qquad
\\ \label{min:Deta}
{\cal B}_{2\alpha } &=& D\eta_{1\alpha} - i \,( 10 E^a \Gamma_{a}
+ i B_1^{ab} \Gamma_{ab})_{\alpha\beta} \wedge \psi^\beta \; .
\qquad
\end{eqnarray}
 The expression for $A_3$ then reads
\begin{eqnarray}
\label{A3=Beta} A_3 \! &=& {1\over 4!} B_1^{ab} \wedge E_a \wedge
E_b \; - {1\over 3^.5!} B_{1\, ab} \wedge B_1^a{}_c \wedge
B_1^{cb} +
 {i \over 5! \, 4 }  \psi^\beta \wedge \eta^\alpha \wedge
\left( 10 \, E^a \Gamma_{a\alpha\beta} + i \, B_1^{ab}
\Gamma_{ab\;\alpha\beta} \right) \; . \qquad \hspace{-1cm} \nonumber \\
{} \hspace{-1cm}
\end{eqnarray}
Its curvature is expressed by (as above, we set
${\mathbf{R}}^a=0$, which is valid both for a description in standard
superspace and for the component approach)
\begin{eqnarray}
\label{F4=Beta} \mathbf{R}_4 &=&  {1\over 4!} {\cal B}_2^{ab}
\wedge E_a \wedge E_b \; + {1\over 5!} {\cal B}_2^{ab} \wedge
B_{1\; bc} \wedge B_1^{c}{}_{a} +
 {i \over 5! \, 4 }  {\mathbf{R}}^\beta \wedge \eta_1^\alpha
\wedge \left( 10 \, E^a \Gamma_{a\alpha\beta} + i \, B_1^{ab}
\Gamma_{ab\;\alpha\beta} \right) -
\nonumber \\
& - & {i \over 5! \, 4 }  \psi^\beta \wedge {\cal B}_2^\alpha
\wedge \left( 10 \, E^a \Gamma_{a\alpha\beta} + i \, B_1^{ab}
\Gamma_{ab\;\alpha\beta} \right) - {1 \over 5! \, 4 }  \psi^\beta
\wedge \eta^\alpha_1 \wedge {\cal B}_2^{ab}
\Gamma_{ab\;\alpha\beta}\; .
\end{eqnarray}
 These relatively simple expressions for $A_3$ and $\mathbf{R}_4 $ will be
 useful, in particular, for an analysis of the supergravity action with
composite $A_3$ in Sec. 5.

For vanishing curvatures Eqs. (\ref{min:Ta=})--(\ref{min:Deta})
become the MC equations for the generalized supersymmetry algebra
$\tilde{\mathfrak{E}}_{min}=\tilde{\mathfrak{E}}^{(66|64)}$
associated to the superspace group $\Sigma^{(66|64)}$ with
coordinates
\begin{eqnarray}\label{66-64}
\tilde{\Sigma}_{min}= \tilde{\Sigma}^{(66|64)}\; : \qquad (x^\mu ,
y^{\mu\nu} , \theta^{\alpha}, \theta^\prime_\alpha )\; .
\end{eqnarray}
Explicitly, these  MC forms are ($\omega^{ab}=0$)
\begin{eqnarray}\label{66-Ea}
  E^a &:=& \Pi^\mu \delta_\mu{}^a =  dx^\mu \delta_\mu{}^a -
 i d\theta^{\alpha}
\Gamma^a_{\alpha\beta} \theta^\beta \; ,
\\ \label{66-Ef}
\psi^\alpha &:=& \Pi^\alpha =  d\theta^{\alpha} \; ,
\\ \label{66-Eb2}
B_1^{ab} &:=&  \Pi^{ab}= dy^{\mu\nu}\delta_\mu{}^a\delta_\nu{}^b -
d\theta^{\alpha}
\Gamma^{ab}_{\alpha\beta} \theta^\beta \; ,
\qquad
\\ \label{66-Ef2}
\eta_{1\alpha} &:= & \Pi^\prime_\alpha = d\theta^{\prime}_\alpha +
i \left(10\, \Pi^a\Gamma_a + i \Pi^{ab}
\Gamma_{ab}\right)_{\alpha\beta} \theta^\beta  - {20\over 3} \,
d\theta\Gamma^a\theta \, (\Gamma_a\theta)_\alpha - {2\over 3}
d\theta\Gamma^{ab}\theta \, (\Gamma_{ab}\theta)_\alpha \; .
\nonumber \\ {}
\end{eqnarray}

\setcounter{equation}0
\section{On the $D=11$ supergravity action with composite $A_3$}

To analyze   possible dynamical consequences of a composite
structure for $A_3$  let us follow the proposal in \cite{D'A+F}
and consider the first order supergravity action with
 a composite $A_3$.

\subsection{
Equations of motion, Noether identities and extra gauge
symmetries.}

The equations of motion for the `free' standard CJS supergravity
(see Sec. 2.4 for a brief review)  include Eq. (\ref{eqmA3=}),
\begin{eqnarray}\label{cG=0}
{\delta S \over \delta A_3} := {\cal G}_8 =0 \; .
\end{eqnarray}
We ask ourselves what would be the consequences of a composite
structure of $A_3$ field.

Our minimal solution given by the Eqs.
 (\ref{A3=Beta}), (\ref{F4=Beta}), (\ref{min:Deta}) (see (\ref{minimal}))  allows for a
simple discussion of this problem; this will also exhibit some
properties relevant for the generic solution, Eqs. (\ref{A3=Ans}),
(\ref{sEq=g(s)}).

To this aim one may just insert the expression   (\ref{A3=Beta}),
(\ref{F4=Beta}) (or (\ref{A3=Ans}) with any allowed set of constants
(\ref{sEq=g(s)})) into the first order action (\ref{S11:=}), without
assuming the FDA relations (\ref{min:Ta=})--(\ref{min:Deta}) [or
(\ref{CJS:Ta=})--(\ref{CJS:RL=}), (\ref{cBab})--(\ref{cDeta}) with
(\ref{s-def})] from the start; their r\^ole and re--appearance will
be discussed below.

As noticed in \cite{D'A+F}, the action with such an insertion
would be very large and hard to handle. To overcome this
difficulty we shall deal with the standard supergravity action,
but with the understanding  that  $A_3$ is made out of the new and
old fields and given by Eq. (\ref{A3=Beta}) or by Eqs.
(\ref{A3=Ans}), (\ref{sEq=g(s)}), so that its variation is not
independent.

Let us begin by the minimal solution. As it follows from Eq.
(\ref{A3=Beta}),
\begin{eqnarray}
\label{varA3} \delta A_3  & = & {1\over 4!} E_a \wedge E_b \wedge
\delta B_1^{ab} \, - {1\over 5!} B_{1\,a}{}^c \wedge B_{1\,cb}
\wedge \delta B_1^{ab} - {1 \over 5! \, 4 } \psi^\beta \wedge
\eta^\alpha_1 \Gamma_{ab\;\beta\alpha} \wedge {\delta B}_1^{ab}
\nonumber \\ &-&
 {i \over 5! \, 4 }  \psi^\beta \wedge ( 10 \, E^a
\Gamma_{a\alpha\beta} + i B_1^{ab} \Gamma_{ab\;\alpha\beta} )
\wedge \delta \eta_1^\alpha   \, ,
\end{eqnarray}
where we have neglected the terms with the variation of the
graviton and the gravitino (which would give contributions
proportional to ${\cal G}_8$ in the Einstein and Rarita-Schwinger
equations of supergravity).

The variation of the supergravity action $S$ with respect to
$B_1^{ab}$ thus reads
\begin{eqnarray}\label{varB}
&& {\delta S \over \delta B_{1\, ab}} =  {\delta S \over \delta
A_3} \wedge {\delta A_3 \over \delta B_{1\; ab} } = {1\over 4!}
{\cal G}_8 \wedge \left(E^a\wedge E^b - {1\over 5} B_1^{ac}\wedge
 B_{1\,c}{}^b - {1\over 20}
\psi \wedge \eta \, \Gamma_{ab}\right)\; . \nonumber \\
\end{eqnarray}

The first order action $S$ from \cite{D'A+F} is the integral over
$D$=11 spacetime $M^{11}$ or an eleven-dimensional bosonic surface
${\cal M}^{11}$ in the standard superspace (or even in a group
manifold \cite{D'A+F,rheoB}). For simplicity we will consider
first in this section the $M^{11}$ case (Eq. (\ref{S11:=})); the
case of the rheonomic action (Eq. (\ref{S11cM})) will be considered in
Sec. 5.3. The vielbein forms $E^a$ provide a basis to express
forms on $M^{11}$. This implies that Eq. (\ref{varB}) has the
expression
\begin{eqnarray}\label{varBK}
{\delta S \over \delta B_{1\, ab}}=
 {1\over 4!} {\cal G}_8 \wedge E^c\wedge E^d \; {\cal K}_{cd}{}^{ab}\; , \quad
\end{eqnarray}
where the matrix
\begin{eqnarray}\label{K=()}
 {\cal K}_{cd}{}^{ab} =
\delta_{[c}{}^a \delta_{d]}^{b} + {1\over 5}
B_{[c}{}^{ae}B_{d]}{}^{bf}\eta_{ef} - {1\over 20}
\psi_{[c}{}^\beta \, \eta_{d]}{}^\alpha \,
\Gamma_{\alpha\beta}^{ab}\; \quad
\end{eqnarray}
may be quite generally assumed to be invertible. Indeed,  one may
think of {\it e.g.}, weak $B_c^{ab}$ fields, in which case the
second term is small and the third nilpotent. Then one may state
that
\begin{eqnarray}\label{varBsim}
 &  det({\cal K}_{ab}{}^{cd} )\not=0   &
\nonumber \\
{\delta S \over \delta B_1^{ab}}=  0 & \Longrightarrow &
 {\cal G}_8 \wedge E^c\wedge E^d =0 \; . \qquad
\end{eqnarray}
 The last equation clearly implies the standard equations of
motion, Eq. (\ref{cG=0}), but now for a composite, rather than
fundamental  $A_3$. Thus one may state, at least within the
$det({\cal K}_{ab}{}^{cd} )\not=0$ assumption, that the variation
with respect to the $B_1^{ab}$ field produces the same equations
as the variation with respect to the CJS three-form $A_3$,
\begin{eqnarray}\label{varB=varA}
&  det({\cal K}_{ab}{}^{cd} )\not=0  \!\!\!  &
\nonumber \\
{\delta S \over \delta B_1^{ab}}=  0 \qquad & \Longrightarrow &
 {\cal G}_8 := {\delta S \over \delta A_3} =0 \; .  \qquad
\end{eqnarray}

A few remarks are appropriate at this stage. The first is that the
above considerations, simplified by the use of the minimal solution
in Eq. (\ref{A3=Beta}), can be extended to the general case, which
has a more complicated expression for $A_3$ that includes the
$B_1^{a_1\ldots a_5}$ field. The second one is that, considered on
the $D$=11 spacetime, $B_1^{ab}=E^cB_c^{ab}(x)$ involves a three
index tensor $B_c^{ab}(x)=-B_c^{ba}(x)$ with reducible symmetry
properties (product of two Young tableaux),
\begin{eqnarray}\label{B=Yt}
B_{c\; ab} \; \sim \;  {}^{\fbox{}} \otimes {}^{\fbox{}}_{\fbox{}}
= {}^{\fbox{}\fbox{}}_{\fbox{}} \oplus {\begin{matrix}\fbox{} \cr
\vspace{-0.6cm} \cr \fbox{} \cr  \vspace{-0.6cm} \cr \fbox{}
\end{matrix}}\qquad ,
\end{eqnarray}
 and thus
carries more degrees of freedom than $A_3=1/3! E^c\wedge E^b\wedge
E^a A_{abc}(x)$ does since $A_{abc}(x)$ is  fully antisymmetric
$A_{abc}=A_{[abc]}$,
\begin{eqnarray}\label{A=Yt}
A_{abc}\; \sim  \; {\begin{matrix}\fbox{}\cr  \vspace{-0.55cm} \cr
\fbox{}\cr  \vspace{-0.55cm} \cr \fbox{}\end{matrix}}\qquad .
\end{eqnarray}
Then, as a variation with respect to $B_1^{ab}$ produces (for
$det({\cal K}_{[ab]}{}^{[cd]} )\not=0$) the same equations as the
variation with respect to $A_3$, one concludes that the action for a
composite $A_3$ must possess local symmetries that make the {\it
extra} ({\it i.e} $\;{}^{\fbox{}\fbox{}}_{\fbox{}}$ but {\it not}
$\begin{matrix}\fbox{} \cr \vspace{-0.7cm} \cr \fbox{} \cr
\vspace{-0.7cm} \cr \fbox{} \end{matrix}\; )$ degrees of freedom in
$B_1^{ab}$ pure gauge. Similarly, one may expect to have an extra
local fermionic symmetry under which the new fermionic fields
$\eta_{a\alpha}(x)$ in $\eta_{1\alpha} =E^a\eta_{a\alpha}(x)$ are
also pure gauge. In the case of a more general solution and
accordingly a more complicated expression for $A_3$, one also
expects a gauge symmetry that makes the five-index one--form fields
in $B_1^{a_1\ldots a_5}=E^bB_b^{a_1\ldots a_5}(x)$ pure gauge.

This is indeed the case. Actually the fact that the above $\delta
B_1^{ab}= E^c \delta B_{c}{}^{ab} $ variation produces the same
result as the variation with respect to $\delta A_{abc}= \delta
A_{[abc]}$ (see Eqs. (\ref{varBsim}) and (\ref{cG=0})) plays the
r\^{o}le of Noether identities for all these `extra' gauge
symmetries. Let us show, for instance, that the supergravity action
with $A_3$ with the simple composite structure of
Eq.~(\ref{A3=Beta}) does possess extra fermionic gauge symmetries
with a spinorial one-form parameter. Indeed, the equations of motion
for $\eta_{1\, \alpha}$,
\begin{eqnarray}\label{vareta}
{\delta S \over \delta \eta_{1}^{\alpha} } = 0\; \quad \Rightarrow
\qquad {\cal G}_8 \wedge \psi^\beta \wedge \left( 10 \, E^a
\Gamma_{a\alpha\beta} + i \, B_1^{ab} \Gamma_{ab\;\alpha\beta}
\right) =0 \; ,
\end{eqnarray}
are satisfied identically on the $B_1^{ab}$  equations of motion
(${\cal G}_8=0$ for $det({\cal K}_{[ab]}{}^{[cd]} )\not=0$, Eqs.
(\ref{varBsim})). This is a Noether identity that indicates the
presence of a local fermionic symmetry  with  parameter
$\chi_{1\alpha}$, $\chi_{1\alpha}= E^a \chi_{a\alpha}$, such that
\begin{eqnarray}\label{dchi}
\delta_{\chi}\eta_{1\alpha} &=& \chi_{1\alpha} \; , \quad  \\
\label{dchiB} \delta_{\chi}B_1^{ab} &=& {i\over 16} {\cal
K}^{-1}{}^{[ab]}{}^{[cd]} \, \psi_c{}^\alpha (10 \Gamma_d +
iB_d{}^{ef}\Gamma_{ef})_{\alpha\beta} \,
\chi_1^\beta \; . \quad \nonumber \\
\end{eqnarray}
We can see that the transformations (\ref{dchi}), (\ref{dchiB})
leave invariant the composite three--form $A_3$ (\ref{A3=Beta})
considered as a form on spacetime. If the form $A_3$ in
(\ref{A3=Beta}) is now considered as defined on standard
superspace $\Sigma$  or on a larger supermanifold
$\tilde{\Sigma}$, the $A_3$ on ${\cal M}^{11}\subset {\Sigma}$ or
$\tilde{\Sigma}$ is still preserved by (\ref{dchi}) for
$\chi_{a\alpha}(\tilde{Z}(x))$ with a more complicated expression
for $\delta_{\chi}B_1^{ab}(\tilde{Z}(x))$.

In the same way, having in mind that the contribution of {\it any}
variation of the fundamental fields in $\delta A_3$ on ${M}^{11}$
is always given by an {\it antisymmetric} third rank tensor
contribution, one concludes that {\it any} contribution to $\delta
A_3$ from an arbitrary variation of the
${}^{\fbox{}\fbox{}}_{\fbox{}}$ irreducible part of $\delta
B_c{}^{ab}$ (which is also an antisymmetric {\it contribution})
can always  be compensated by a contribution of a proper
transformation of its completely antisymmetric part $\delta
B_{[cba]}$, ${\begin{matrix}\fbox{}\cr \vspace{-0.6cm} \cr
\fbox{}\cr \vspace{-0.6cm} \cr \fbox{}\end{matrix}}\quad$.

When the more general form for $A_3$, (Eqs. (\ref{A3=Ans}),
(\ref{sEq=g})) is considered, the same reasoning shows that any
transformations of the new form $B_1^{a_1\ldots a_5}$ can be
compensated by some properly chosen $B_1^{ab}$ transformations.
The key point is that the coefficient $\lambda$ in (\ref{sEq=g})
never vanishes. Hence (omitting $\delta E^a$ and $\delta
\psi^\alpha$)
\begin{eqnarray}
\label{varA3g} \delta A_3 &=& - {\lambda \over 4} E^c \wedge E^d
\wedge {\cal K}_{cd}{}^{ab} \delta B_{1\; ab}  \; + {\cal S}_{2
a_1\ldots a_5} \wedge \delta B_1^{a_1\ldots a_5} +  {\cal
S}_{2}^{\alpha} \wedge \delta \eta_{1\, \alpha} \; =
\\ \label{varA=l}
 & = & -   {\lambda \over 4} E^a \wedge E^b \wedge E^c \,
\delta B_{[c\; ab]}  + {\cal O}( B_1 \wedge B_1) +  {\cal O}(
\psi_1 \wedge \eta_1) \; ,
\end{eqnarray}
\begin{eqnarray} \label{K=g} {\cal K}_{cd}{}^{ab} &=
\delta_{[c}{}^{a}\delta_{d]}{}^{b} + {\cal O}( B_1 \wedge B_1) +
{\cal O}( \psi_1 \wedge \eta_1)  \; , \qquad
\\
\nonumber & \lambda = {(20\gamma^2_1 +\delta^2)\over 5
 (2 \gamma_1-  \delta)^2}  \equiv {1 \over 5} \; \frac{s^2+2s+6}{s^2} \; \not=0
\end{eqnarray}
and  the variation of the completely antisymmetric part
$B_{[abc]}$ of $B_1^{ab}=E^cB_c{}^{ab}$ always reproduces (for an
invertible ${\cal K}$ (\ref{K=g})) the same equation ${\cal
G}_8=0$ as it would an independent, fundamental  three-form
$A_3$.

\subsection{Free differential algebra for the `new'
fields}

One may ask at what stage  the FDA relations (\ref{cBab}),
(\ref{cB15}), (\ref{cDeta}) appear when the first order supergravity
action \cite{D'A+F,J+S99} with a composite $A_3$ field [Eqs.
(\ref{A3=Ans}), (\ref{sEq=g})] is considered and whether there are
any conditions on the new curvatures, as Eqs. (\ref{fdTa=}),
(\ref{fdF4=}) for $\mathbf{R}^a$ (Eq. (\ref{CJS:Ta=})) and
$\mathbf{R}_4$ (Eq.(\ref{CJS:R4=})). Let us recall that in the first
order action the latter equations are not imposed by hand,  but
appear as equations of motion. The action (\ref{S11:=}) of refs.
\cite{D'A+F,J+S99} include the auxiliary field $F_{abcd}$ and the
variation with respect to it produces Eq. (\ref{cdA-=0}), which is
equivalent to the FDA relation (\ref{CJS:R4=}) with (\ref{fdF4=}).
The variation with respect to an independent spin connection
$\omega^{ab}$ produces Eq. (\ref{varS=Ta}) which is equivalent to
the FDA relation (\ref{CJS:Ta=}) with (\ref{fdTa=}).

As it was shown in Sec. 4, Eqs. (\ref{CJS:R4=}) can be solved by
expressing $A_3$ in terms of the one--forms $E^a$, $B_1^{ab}$,
$B_1^{a_1\ldots a_5}$, $\psi^\alpha$, $\eta_{1\, \alpha}$ by Eq.
(\ref{A3=Ans}) for any set of constants given by Eqs.
(\ref{sEq=g}) (or (\ref{sEq=g(s)})), provided the one-forms
satisfy the FDA (\ref{CJS:Ta=}), (\ref{CJS:Tf=}), (\ref{CJS:RL=})
and (\ref{cBab}), (\ref{cB15}), (\ref{cDeta}) with the same
constants $\delta, \gamma_1, \gamma_2$ (with the same $s$). Eq.
(\ref{A3=Ans}) implies also the expression for $\mathbf{R}_4$
through the field strengths of the one--form fields (two--form
curvatures of the soft FDA algebra) (see Eq. (\ref{R4=Ans})).

With this in mind, studying the first order supergravity action
which produces Eqs.  (\ref{cdA-=0}) and (\ref{varS=Ta}),
 one may just use the FDA relations (\ref{cBab}), (\ref{cB15}),
(\ref{cDeta}) (with $\delta, \gamma_1, \gamma_2$ expressed through
the same parameter $s$ as in (\ref{sEq=g(s)}) used to construct
$A_3$) to substitute
 the
expressions
\begin{eqnarray}
\label{dB2=cB2} DB_1^{ab} &=& - \psi^\alpha  \wedge \psi^\beta \,
\Gamma^{ab}_{\alpha\beta} + {\cal B}_2^{ab} \; , \qquad
\\
\label{dB5=cB5} DB_1^{a_1\ldots a_5}  &=& - i \psi^\alpha  \wedge
\psi^\beta \,
\Gamma^{a_1\ldots a_5}_{\alpha\beta} + {\cal B}_2^{a_1\ldots a_5} \; , \qquad \\
 \label{det=cBf}
D\eta_{1\alpha} &=& + i \, \delta \, E^a \wedge \psi^\beta
\Gamma_{a\, \alpha\beta} + \gamma_1 \, B_1^{ab} \wedge \psi^\beta
\Gamma_{ab\, \alpha\beta}   + i \, \gamma_2 \, B_1^{a_1\ldots a_5}
\wedge \psi^\beta \Gamma_{a_1\ldots a_5\alpha\beta} + {\cal
B}_{2\alpha } \; ,  \qquad \nonumber \\ {}
\end{eqnarray}
for $DB^{ab}_1$, $DB_1^{a_1\ldots a_5}$, $D\eta_{1\alpha}$ in $dA_3$
of (\ref{cdA-=0}). What one gains making just these substitutions
in Eq. (\ref{cdA-=0}) is that all the terms without curvatures
coming from the first term ($dA_3(B_1^{ab}, \ldots )$) are
cancelled by the second term ($a_4$ of Eq. (\ref{a4:=})). Thus,
the only consequences of Eq. (\ref{cdA-=0}) would be an expression
for $F_4$ through the newly defined field curvatures and
$\mathbf{R}^\alpha$ (Eq. (\ref{CJS:Tf=}))
\begin{eqnarray}\label{eeeeF}
E^a\wedge E^b \wedge E^c \wedge E^d\, F_{abcd} & = & \lambda {\cal
B}_2^{ab} \wedge E_a \wedge E_b + \ldots - \nonumber \\ && - 2i
\psi^\beta \wedge \eta_1^\alpha \wedge \left(-i  \beta_2 \, {\cal
B}_2^{ab} \Gamma_{ab} + \beta_3 \, {\cal B}_2^{abcde}
\Gamma_{abcde} \right)_{\alpha\beta} \;
\end{eqnarray}
(see Eq.(\ref{R4=Ans}) for the full expression  of the
right--hand--side).

\bigskip

 As we discussed in Sec. 5.1., the variation of the new fields
produces the only nontrivial equation, Eq. (\ref{varB=varA}),
which formally coincides with the original three--form equation
(\ref{cG=0}), but now involving the composite $A_3$ and its field
strength  $F_{abcd}$ [see Eq. (\ref{cdA-=0}) which appears as a
result of varying  $F_{abcd}$ in the action (\ref{S11:=}),
(\ref{L11:=}), independently of whether $A_3$ is composite or
fundamental]. This reflects the existence of the extra gauge
symmetries (see Sec. 5.1) that make that the theory with a
composite $A_3$ carries the same number of degrees of freedom as
the original CJS supergravity with a fundamental $A_3$. What we
have found now is that, besides of this, Eq. (\ref{eeeeF}) is the
only relation imposed on the new field strengths ${\cal
B}_2^{ab}$, $ {\cal B}_2^{a_1\ldots a_5}$, ${\cal B}_{2\,\alpha}$
by the first--order $D=11$ supergravity action (\ref{S11:=}),
(\ref{L11:=}) with a composite $A_3$. This also reflects the
existence of the extra gauge symmetries, as these  make the
detailed properties of the curvatures (${\cal B}^{abcde}_2$, $
{\cal B}_2^{a_1\ldots a_5}$, ${\cal B}_{2\,\alpha}$)  of the
additional gauge fields inessential; their only relevant
properties are that the field strength $F_{abcd}$ is constructed
out of them in accordance with Eq. (\ref{eeeeF}) and that such a
composite field strength obeys Eq. (\ref{cG=0}).

Thus, on the one hand,  the underlying gauge group structure implied
by the new one--form fields allows us to treat the $D=11$
supergravity as a gauge theory of the
 $\tilde{\Sigma}(s\not=0)\times\!\!\!\!\!\!\!\supset SO(1,10)$ supergroup,
 that replaces  the superPoincar\'e one. On the other hand, the supergravity
 action (\ref{S11:=}), (\ref{L11:=}) with a composite $A_3$ also possesses
`extra' gauge symmetries  (not in
$\tilde{\Sigma}(s\not=0)\times\!\!\!\!\!\!\!\supset SO(1,10)$)
 that the {\it additional} degrees of freedom in the
`new' fields $B_1^{ab}$, $B_1^{a_1\ldots a_5}$, $\eta_{1\alpha}$
pure gauge ({\it i.e.} $B_1^{ab}$, $B_1^{a_1\ldots a_5}$,
$\eta_{1\alpha}$ carry the same number of physical degrees of
freedom as the fundamental $A_3$ field). In this sense the geometric
$\tilde{\Sigma}(s\not=0) \times\!\!\!\!\!\!\!\supset SO(1,10)$
symmetry, although manifest, gives only a part of the gauge
symmetries of the  supergravity action (\ref{S11:=}), (\ref{L11:=})
with a composite $A_3$.

One might conjecture that the superfluous degrees of freedom in the
`new' one--form fields,  which are pure gauge in the pure
supergravity action, would become `alive' when supergravity is
coupled to some M--theory objects. These could not be the usual
M--branes as they couple to the standard fields and, hence, all the
gauge symmetries preserving the composite $A_3$ would remain
preserved. Thus one might think of coupling of supergravity through
some new action containing explicitly the new one--forms. A guide in
the search for such an action would be the preservation of the gauge
symmetries of the underlying
 $\tilde{\Sigma}(s\not=0)\times\!\!\!\!\!\!\supset SO(1,10)$
 gauge supergroup.

\subsection{Composite $A_3$ in the rheonomic action. A possible
way to enlarged superspace}

All the above discussion on the `extra' gauge symmetries (Sec.
5.1) applies also for the rheonomic action (\ref{S11cM}) with
${\cal M}^{11}$ being an arbitrary surface in superspace. In
short, this follows from the fact that all the one--forms on such
a surface can be decomposed using the basis provided by the
pull--back $E^a(\tilde{Z}(x))=d\tilde{Z}^M(x)
E_M{}^a(\tilde{Z}(x))= dx^\mu
\partial_\mu \tilde{Z}^M(x) E_M{}^a(\tilde{Z}(x))$ of the bosonic
supervielbein $E^a(Z)=dZ^M E_M{}^a(Z)$.

Indeed, in the matrix  $\partial_\mu \tilde{Z}^M(x)
E_M{}^a(\tilde{Z}(x))= E_\mu{}^a(\tilde{Z}(x)) + \partial_\mu
\tilde{\theta}^{\check{\alpha}}(x)
E_{\check{\alpha}}{}^a(\tilde{Z}(x))$ the first term is given by
an invertible matrix $E_\mu{}^a$ while the second is nilpotent.
Hence there exists a matrix $\mathrm{E}_a{}^{\mu}=
\mathrm{E}_a{}^{\mu} (\tilde{Z}, \partial_\nu \tilde{Z})$ such
that $\mathrm{E}_a{}^{\mu} [E_\mu{}^b(\tilde{Z}(x)) +
\partial_\mu \tilde{\theta}^{\check{\alpha}}(x) E_{\check{\alpha}}{}^b(\tilde{Z}(x))]=
\delta_a{}^b$.  This is tantamount to saying that $dx^\mu =
E^a(\tilde{Z}(x)) \mathrm{E}_a{}^{\mu}$. Using this we may express
a superspace differential form on ${\cal M}^{11}$ in the
$E^a(\tilde{Z}(x))$ basis. In particular,  the superform
$B_1^{ab}(Z)=dZ^M B_M^{ab}(Z)$ on ${\cal M}^{11}$, $\;
B_1^{ab}(\tilde{Z}(x))=d\tilde{Z}^M(x) B_M^{ab}(\tilde{Z}(x)) =
dx^\mu \partial_\mu \tilde{Z}^M(x) B_M^{ab}(\tilde{Z}(x))$, may be
written as ${B}_1^{ab}(\tilde{Z}(x))= {E}^c(\tilde{Z}(x))
\tilde{B}_c{}^{ab}$. With this in mind the above considerations on
local symmetries may be extended to the case of  superforms on
arbitrary eleven-dimensional bosonic surfaces.

The new aspect that the composite structure of $A_3$ brings to the
rheonomic action is that the surface ${\cal M}^{11}$ is now
allowed to be an arbitrary one {\it in the enlarged superspace}
$\Sigma^{(528|32+32)}(s\neq 0)$ with coordinates ${\cal Z}^{{\cal
N}}:= ({y}^{\nu}, y^{\nu_1\nu_2}, {y}^{\nu_1\ldots \nu_5},
{\theta}^{\check{\alpha}},
\tilde{\theta}^\prime_{\check{\alpha}})$. With the identification
$y^\mu=x^\mu$, such a surface is defined by its set of embedding
functions, namely,  the (already familiar)
$\tilde{\theta}^{\check{\alpha}}(x)$ plus
$\tilde{y}^{\mu_1\mu_2}(x)$, $\tilde{y}^{\mu_1\ldots \mu_5}(x)$,
and $\tilde{\theta}^\prime_{\check{\alpha}}(x)$. More generally,
one may define $x^\mu$ to be local coordinates of ${\cal M}^{11}$
and  distinguish them from the corresponding bosonic coordinates
$y^\mu$ of $\tilde{\Sigma}^{(528|32+32)}(s)$ to define ${\cal
M}^{11}$ parametrically as
\begin{eqnarray}\label{calMinS}
{\cal M}^{11} & \subset & \tilde{\Sigma}^{(528|32+32)}(s) \;  :
\qquad {\cal Z}^{{\cal N}} = \tilde{{\cal Z}}^{{\cal N}}(x) \qquad
\\ \nonumber \Leftrightarrow &&  \left(\begin{matrix} y^\mu=\tilde{y}^\mu(x)  \cr
y^{\mu_1\mu_2}=\tilde{y}^{\mu_1\mu_2}(x) \cr y^{\mu_1\ldots
\mu_5}=\tilde{y}^{\mu_1\ldots \mu_5}(x) \cr
{\theta}^{\check{\alpha}}= \tilde{\theta}^{\check{\alpha}}(x) \cr
{\theta}^\prime_{\check{\alpha}}=
\tilde{\theta}^\prime_{\check{\alpha}}(x)
\end{matrix}\right) \qquad .
\end{eqnarray}

 The standard $D$=11 rheonomic action $S_{11}^{rh}$ in
 Eq. (\ref{S11cM}) with a fundamental $A_3$ form
 may also be considered as an action for a spacetime filling
10--brane ($p=D-1$=10) in standard superspace. However, some
properties of this action are quite unusual for the familar
$p$--branes. These include the symmetry under arbitrary changes of
the surface ${\cal M}^{11}$ that allows us to gauge all the
fermionic coordinate functions $\tilde{\theta}^{\check{\alpha}}(x)$
away to obtain the standard spacetime (but first order) supergravity
action, the local supersymmetry with a number of parameters equal to
the number of fermionic coordinate functions, and the fact that this
latter symmetry still is present when the
$\tilde{\theta}^{\check{\alpha}}(x)$ are gauged away (see
\cite{Regge,rheoB} and \cite{BAIL01} for further discussion).

For a composite $A_3$, the rheonomic action can be written as an
integral over a surface ${\cal M}^{11}$ in the enlarged superspace
$\tilde{\Sigma}^{(528|32+32)}(s)$, Eq. (\ref{calMinS}),
\begin{eqnarray}
\label{S11cMS} \tilde{S}_{11}^{rh} & = & \int_{{\cal M}^{11}\in
\tilde{\Sigma}(s)} {\cal L}_{11} ({\cal Z}^{{\cal N}}) =
\int_{{M}^{11}} {\cal L}_{11} (\tilde{{\cal Z}}^{{\cal N}}(x)) \;
. \qquad
\end{eqnarray}
This looks like an action of a brane which is no longer spacetime
filling, as the dimension of the bosonic body of
$\tilde{\Sigma}^{(528|32+32)}(s)$ superspace is $528$.  The
functional (\ref{S11cMS}) is of course  the rheonomic action for
supergravity which, due to the  extra gauge symmetries, is
equivalent to a spacetime (component) first order action but with
a composite $A_3$ field. However, the fact that ${\cal L}_{11}
({\cal Z}^{{\cal N}})$ can be looked at as a form on the extended
superspace $\tilde{\Sigma}^{(528|32+32)}(s)$ suggests trying to
search for an embedding of $D=11$ supergravity in a theory defined
on $\tilde{\Sigma}^{(528|32+32)}(s)$. In particular, it is
tempting to look for possible 10--brane models in
$\tilde{\Sigma}^{(528|32+32)}(s)$. For instance, one might search
for a superembedding condition (see \cite{bpstv,Dima}) of the
standard ${\Sigma}^{(11|32)}$ superspace (the worldvolume
superspace of such a hypothetical brane) into
$\tilde{\Sigma}^{(528|32+32)}(s)$ that could reproduce the
on-shell eleven--dimensional supergravity constraints. Such a
study is, however, beyond the scope of this paper.

\setcounter{equation}0
\section{Conclusions and outlook}

We have studied here the consequences of a possible composite
structure of the three--form field $A_3$ of the standard CJS
$D=11$ supergravity. In particular, we have provided the
derivation of our previous result \cite{Lett} by which the $A_3$
three-form field may be expressed in terms of the one--form gauge
fields $B_1^{ab}$, $B_1^{a_1\ldots a_5}$, $\eta_{1\alpha}$, $E^a$,
$\psi^\alpha$ associated with  a {\it family} of superalgebras
$\tilde {\mathfrak E}(s)$, $s\not=0$, corresponding to the
supergroups $\tilde{\Sigma}(s)=\tilde{\Sigma}^{(528|32+32)}(s)$.
Two values of the $s$ parameter recover the two earlier
D'Auria--Fr\'e \cite{D'A+F} decompositions of $A_3$, while one
value of $s$ leads to a simple expression for $A_3$ that does not
involve
 $B^{a_1\ldots a_5}_1$. The supergroups
 $\tilde{\Sigma}(s) \times\!\!\!\!\!\!\supset SO(1,10)$ with $s\not=0$
 may be considered as nontrivial deformations of the
 $\tilde{\Sigma}(0) \times\!\!\!\!\!\!\supset SO(1,10)\subset
 \tilde{\Sigma}(0) \times\!\!\!\!\!\!\supset Sp(32)$ supergroup,
which is itself the expansion \cite{Lett,JdA02} $OSp(1|32)(2,3,2)$
of $OSp(1|32)$. For any $s\not=0$, $\tilde{\Sigma}(s)
\times\!\!\!\!\!\!\supset SO(1,10)$ may be looked at as a hidden
gauge symmetry of the $D=11$ CJS supergravity generalizing the
$D$=11 superPoincar\'e group
 $\Sigma^{(11|32)}\times\!\!\!\!\!\!\supset SO(1,10)$.

We have stressed the equivalence between the problem of searching
for a composite structure of the $A_3$ field and, hence, for a
hidden gauge symmetry of $D=11$ supergravity, and that  of
trivializing a four--cocycle of the standard $D=11$ supersymmetry
algebra $\mathfrak{E}^{(11|32)}$
 on the enlarged superalgebras
 $\tilde{\mathfrak{E}}^{(528|32+32)}(s)$, $s\neq 0$.
The generators of $\tilde{\mathfrak{E}}^{(528|32+32)}(s)$  are in
one--to--one correspondence with the one--form  fields $E^a$,
$\psi^\alpha$, $B_1^{ab}$, $B_1^{a_1\ldots a_5}$, $\eta_{1
\alpha}$. For zero curvatures these fields can be identified with
the ${\tilde{\Sigma}}^{(528|32+32)}(s)$--invariant Maurer--Cartan
forms of $\tilde{\mathfrak{E}}^{(528|32+32)}(s)$ which, before
pulling them back to a bosonic eleven--dimensional surface, are
expressed through the coordinates $(x^\mu,
\theta^{\check{\alpha}}, y^{\mu\nu}, y^{\mu_1\ldots \mu_5},
\theta^{\prime}_{\check{\alpha}})$ of the
${\tilde{\Sigma}}^{(528|32+32)}(s)$ superspace.

To study the possible  dynamical consequences of the composite
structure of $A_3$ we have followed D'Auria and Fr\'e original
proposal \cite{D'A+F} of substituting the composite $A_3$ for the
fundamental $A_3$ in the first order CJS supergravity action
\cite{D'A+F,J+S99} (see Sec.~\ref{2.4} for a review). We have
shown that such an action possesses the right number of `extra'
gauge symmetries to make the number of degrees of freedom the same
as in the standard  CJS supergravity. These are clearly symmetries
under the transformations of the new one--form fields that leave
the composite $A_3$ field invariant; their presence is related to
the fact that the new gauge fields enter the supergravity action
only inside the $A_3$ field.

 We would like to mention here some similarities between the
 problem of searching for the composite structure of the $A_3$
 field and the treatment of the Born-Infeld fields of D-branes and
the M5--brane antisymmetric tensor field as composite fields in
\cite{JdA00}. Born-Infeld fields are usually defined as
`fundamental' gauge fields {\it i.e.}, they are given,
respectively, by one-forms $A_1(\xi)$ and a two-form $A_2(\xi)$
directly defined on the worldvolume $\mathcal{W}$. It was shown in
\cite{JdA00} (see also \cite{Saka-98}) that both $A_1(\xi)$ and
$A_2(\xi)$ can be expressed through pull-backs to $\mathcal{W}$ of
forms defined on superspaces $\tilde{\Sigma}$ suitably enlarged by
additional bosonic and fermionic coordinates, in accordance with
the worldvolume fields/extended superspace variables
correspondence principle for super--$p$--branes \cite{JdA00} (see
also \cite{Az-Iz-Mi-04}). The embedding of $\mathcal{W}$ into
$\tilde{\Sigma}$  specifies the dynamics of the composite
$A_1(\xi)$ and $A_2(\xi)$ fields. The extra degrees of freedom
that are introduced by considering $A_1(\xi)$ and $A_2(\xi)$ to be
the pull-backs to $\mathcal{W}$ of forms given on
$\tilde{\Sigma}$, and that produce the composite structure of the
Born--Infeld fields to be used in the superbrane actions, are
removed by the appearance of extra gauge symmetries \cite{JdA00},
as is here the case for the composite $A_3$ field of $D$=11
supergravity. Of course, these two problems are not identical:
 for instance, in the case of $D$=11 supergravity with a composite
 $A_3$, the suitably enlarged {\it flat} superspaces
 $\tilde{\Sigma}(s)=\tilde{\Sigma}^{(528|32+32)}(s)$
solves the associated problem of trivializing the CE cocycle, but
a dynamical $A_3$ field requires `softening'  the
$\tilde{\mathfrak{E}}^{(528|32+32)}(s\neq 0)$ MC equations by
introducing nonvanishing  curvatures; in contrast, the Born-Infeld
worldvolume fields $A_1(\xi)$ and $A_2(\xi)$ are already dynamical
in the flat superspace situation considered in \cite{JdA00}.
Nevertheless, in both these seemingly different situations the
fields/extended superspace variables correspondence leads us to
the convenience of enlarging standard superspace\footnote{We
further note that extended superspaces also appear in the
description \cite{BESE,JdA00} of the strictly invariant
Wess--Zumino terms of the scalar branes. In a similar spirit,
these invariant WZ terms trivialize their characterizing CE
$(p+2)$-cocycles \cite{AT89} on the standard supersymmetry
algebras $\mathfrak{E}^{(D|n)}$, including, of course, that of the
$D=11$ supermem\-brane, since its WZ term is given by the
pull-back to ${\cal W}$ of the three-form potential of the $dA_3$
superspace four-cocycle.}. In this way, all the fields appearing
in the theory (be them on spacetime or on the worldvolume)
correspond to the coordinates of a suitably enlarged superspace.

The above mentioned `extra'  gauge symmetries are also present in
the rheonomic action for $D=11$ supergravity when $A_3$ is a
composite superform. The rheonomic action for a fundamental $A_3$
(shortly reviewed in Sec.~\ref{2.5}) is derived from the spacetime
component first order action just by replacing all the
differential forms on spacetime by superforms on the standard
superspace, pulled back to an eleven--dimensional bosonic  surface
${\cal M}^{11}$. Such a surface is specified by a fermionic
coordinate function $\tilde{\theta}^{\check{\alpha}}(x)$ which is
also considered as a dynamical variable. The rheonomic action
allows one, with the help of additional step of  `lifting' (see
Sec.~\ref{2.5}), to reproduce the standard superspace constraints
of the $D=11$ supergravity (see Sec.~2.2 for a discussion and
\ref{2.3} for their relation with free differential algebras). In
this perspective the composite structure of $A_3$ allow us to
consider ${\cal M}^{11}$ as a surface in an enlarged superspace
$\Sigma^{(528|32+32)}(s)$. As discussed in Sec.~5.3, this  might
indicate a possibility of embedding  $D=11$ supergravity in a more
general theory defined in an enlarged superspace.

To summarize,  the underlying gauge  symmetry $\tilde{\Sigma}(s)
\times\!\!\!\!\!\!\supset SO(1,10)$ of the $D$=11 supergravity is
hidden in the CJS supergravity with a fundamental $A_3$, but
becomes manifest in the action with a composite $A_3$ field.
However, the latter possesses also a set of  {\it extra} gauge
symmetries, due to the fact that the new fields enter the action
inside the composite $A_3$ field only. These extra gauge
symmetries produce that the {\it new} gauge fields, {\it i.e.} the
gauge fields $B_1^{ab}$, $B_1^{a_1\ldots a_5}$, $\eta_{1\alpha}$
corresponding to the coset $\tilde{\Sigma}(s)/\Sigma$, carry the
same number of degrees of freedom as the fundamental $A_3$ field.
In other words, the degrees of freedom in these fields that go
beyond those in the fundamental $A_3$ are pure gauge ones. One may
conjecture that these extra degrees of freedom might be important
in M--theory and that, correspondingly, the extra gauge symmetries
that remove them would be broken by including in the supergravity
action some exotic `matter' terms that couple to the `new'
additional one--form gauge fields. In constructing such an
`M--theoretical matter' action, the preservation of the
$\tilde{\Sigma}(s) \times\!\!\!\!\!\!\supset SO(1,10)$ gauge
symmetry  would provide a guiding principle.

A preliminary study of the local supersymmetry of the new fields
shows  that the preservation of the standard supersymmetry
transformation rules for $A_3$ implies that the pure group
theoretical transformation rules for the
$\tilde{\Sigma}(s)/\Sigma$ gauge fields have to be modified. To
analyze the structure of such a modification one can study the
solution of the Bianchi identities of the $\tilde{\Sigma}(s)
\times\!\!\!\!\!\!\supset SO(1,10)$ gauge FDA, which is tantamount
to studying the  Bianchi identities for the superforms on
$\Sigma^{(528|32+32)}(s)$ superspace.

This brings again  the question of whether $D=11$ supergravity can
be included in a superfield theory defined on the enlarged
superspace $\Sigma^{(528|32+32)}(s)$, $s\neq 0$, in particular in a
hypothetical superfield supergravity in $\Sigma^{(528|32+32)}(s)$. A
generalized supergravity in $\tilde{\Sigma}^{(528|32)}\subset
\tilde{\Sigma}^{(528|32+32)}(s)$ superspace with holonomy group
$GL(32)$ or $SL(32)$ was recently studied \cite{BPST04} (in general
for $\Sigma^{(n(n+1)/2|n)}$, although with emphasis in the
$n=4,8,16$ cases in relation with higher spin theories in
$D=4,6,10$). A study of supergravity in the present
$\tilde{\Sigma}^{(528|32+32)}(s\not=0)$ superspace might lead to a
different result due to the presence of the additional fermionic
variables  and to the natural reduction of the $GL(32)$ structure
group of $\tilde{\Sigma}^{(528|32)}$ down to the $SO(1,10)$
automorphism symmetry of $\tilde{\Sigma}^{(528|32+32)}(s\not=0)$.

\bigskip

We conclude by mentioning that a possible composite structure for
the $A_3$ field has also been considered recently
\cite{Sati03,Moore03} (see also \cite{Lechner+Marchetti03}) in a
different perspective, in connection with the problem of anomalies
in M-theory \cite{Witten:1996} and with M-theory in a
topologically nontrivial situation \cite{HW96,Moore03}. There, the
$A_3$ field is constructed/defined using an  auxiliary
twelve--dimensional $E_8$ gauge theory. It has been asked in
\cite{Moore03} whether the $E_8$ formalism is unique. In this
respect it would be interesting to see whether the composite
structure of $A_3$ field found in \cite{D'A+F}, extended to
$\tilde {\mathfrak E}(s)$ in \cite{Lett}, and developed in the
present paper, could be useful in the context of
\cite{Witten:1996,Moore03}.

\bigskip

{\bf Acknowledgments}. This work has been partially supported by
the research grant BFM2002-03681 from the Ministerio de
Educaci\'on y Ciencia and from EU FEDER funds, by the Generalitat
Valenciana (03/124), by the grant N 383 of the Ukrainian State
Fund for Fundamental Research and the INTAS Research Project N
2000-254. M.P. and O.V. wish to thank the Ministerio de
Educaci\'on y Ciencia and the Generalitat Valenciana,
respectively, for their FPU and FPI research grants. Several
discussions with  J.M. Izquierdo and a conversation with P. van
Nieuwenhuizen are gratefully acknowledged.

\end{document}